\documentclass[aps,twocolumn,pra,floatfix,showpacs,lengthcheck,citeautoscript,superscriptaddress,longbibliography]{revtex4-1}
\usepackage{amsmath}
\usepackage{amssymb}
\usepackage{graphicx}
\usepackage{bm}
\usepackage{hyperref}
\usepackage{color}
\usepackage{times}
\usepackage{MnSymbol}
\usepackage{verbatim}
\usepackage{bbold}

\newcommand{\bea}{\begin{eqnarray}}
\newcommand{\eea}{\end{eqnarray}}
\newcommand{\be}{\begin{equation}}
\newcommand{\ee}{\end{equation}}

\newcommand{\vs}[1]{\hat{\bm{#1}}}

\begin{document}
\title{Majorana fermions on the quantum Hall edge}

\author{Lucila Peralta Gavensky}
\affiliation{Centro At{\'{o}}mico Bariloche and Instituto Balseiro,
Comisi\'on Nacional de Energ\'{\i}a At\'omica (CNEA)--Universidad Nacional de Cuyo (UNCUYO), 8400 Bariloche, Argentina}
\affiliation{Instituto de Nanociencia y Nanotecnolog\'{i}a (INN), Consejo Nacional de Investigaciones Cient\'{\i}ficas y T\'ecnicas (CONICET)--CNEA, 8400 Bariloche, Argentina}

\author{Gonzalo Usaj}
\affiliation{Centro At{\'{o}}mico Bariloche and Instituto Balseiro,
Comisi\'on Nacional de Energ\'{\i}a At\'omica (CNEA)--Universidad Nacional de Cuyo (UNCUYO), 8400 Bariloche, Argentina}
\affiliation{Instituto de Nanociencia y Nanotecnolog\'{i}a (INN), Consejo Nacional de Investigaciones Cient\'{\i}ficas y T\'ecnicas (CONICET)--CNEA, 8400 Bariloche, Argentina}

\author{C. A. Balseiro}
\affiliation{Centro At{\'{o}}mico Bariloche and Instituto Balseiro,
Comisi\'on Nacional de Energ\'{\i}a At\'omica (CNEA)--Universidad Nacional de Cuyo (UNCUYO), 8400 Bariloche, Argentina}
\affiliation{Instituto de Nanociencia y Nanotecnolog\'{i}a (INN), Consejo Nacional de Investigaciones Cient\'{\i}ficas y T\'ecnicas (CONICET)--CNEA, 8400 Bariloche, Argentina}

\begin{abstract}

Superconductivity and the quantum Hall effect are considered to be two cornerstones of condensed matter physics. The realization of hybrid structures where these two effects coexist has recently become an active field of research. In this work, we study a Josephson junction where a central region in the quantum Hall regime is proximitized with superconductors that can be driven to a topological phase with an external Zeeman field. In this regime, the Majorana modes that emerge at the ends of each superconducting lead couple to the chiral quantum Hall edge states. This produces distinguishable features in the Andreev levels and Fraunhofer patterns that could help in detecting not only the topological phase transition but also the spin degree of freedom of these exotic quasiparticles. The current phase relation and the spectral properties of the junction throughout the topological transition are fully described by a numerical tight-binding calculation. In pursuance of the understanding of these results, we develop a low-energy spinful model that captures the main features of the numerical transport simulations in the topological phase.
\end{abstract} 
\maketitle
\section{Introduction}
About 30 years ago, theoretical physicists asked themselves how the Josephson effect would occur between s-wave superconductors coupled to the edge states of a sample in the quantum Hall regime~\citep{Ma1993}. What might have been seen as a bold question, has now become a concrete and tangible possibility~\citep{Mason2016}. Experimental groups have recently managed to make sufficiently transparent contacts between superconductors and quantum Hall states~\citep{Wan2015,Amet2016,Lee2017,Park2017}, not only enabling the measurement of a supercurrent~\citep{Amet2016,Guiducci2018,Seredinski2019}, but also establishing the existence of the so called chiral Andreev edge state~\citep{Zhao2020}, a one-way hybrid electron-hole mode that propagates  along these interfaces~\citep{Hoppe2000}. The electron-hole cyclotron orbits in the semiclassical regime were also recently imaged in a focusing experiment~\citep{Bhandari2020}.

The main physical consequence of the presence of chiral quantum Hall edge states bridging the superconductors in a Josephson junction is that backscattering is ruled out and so conventional Andreev retroreflection is not allowed~\citep{Hoppe2000}. The charge transfer mechanism that produces a supercurrent must then involve the entire perimeter of the Hall bar~\citep{Ma1993,Stone2011,vanOstaay2011,Alavirad2018}, yielding an unusual critical supercurrent $J_c$ as a function of the flux threading the sample. In fact, the current-phase relation is expected to obey a  normal flux quantum $\Phi_0 = hc/e$ periodicity instead of the conventional one with the superconducting quantum $\Phi_0/2$.

In this article, we pose the question of what would happen if the s-wave superconductors were to be replaced by topological ones. In particular, we study the transport and spectral properties of a quantum Hall based junction with one-dimensional superconducting leads that can be driven from a trivial s-wave phase to a p-wave topological phase, where Majorana quasiparticles emerge at the ends of each terminal. We find that this topological phase transition can be detected by analyzing the behavior of the supercurrent in the device, which is entirely carried by the chiral edge channels of the Hall sample. Our main claim is that the Fraunhofer patterns, which describe the modulations of the critical supercurrent as a function of the magnetic field $B_z$ through the quantum Hall region, $J_c(B_z)$, not only reveal the presence of the Majorana fermions, but they also bear information on the spin polarization~\citep{Sticlet2012} of these topologically protected end modes.

The work is organized as follows. In section \ref{II} we introduce the tight-binding model of the Josephson junction. We calculate the supercurrent as a function of the phase difference between the superconducting leads and the critical current profiles as the magnetic flux through the junction is varied in amounts of the order of the flux quantum. We focus on the quantum regime, where only the first Landau level is occupied, and we analyze how these Fraunhofer interference patterns evolve as the leads are  driven from the trivial to the topological phase. The spectral properties of the device are also presented, revealing how the Andreev level spectrum is correlated with the transport simulations. In section \ref{III} we introduce a low-energy spinful model that allows us to reproduce the main features of the full numerical model. We also do a detailed analysis of the limiting case in which the wires behave as spinless p-wave Kitaev chains. In section \ref{IV} we briefly discuss how the transport results are modified when there are two Landau levels occupied in the quantum Hall region. Finally,  we summarize our main results and state some concluding remarks in section \ref{V}.
\section{Tight-binding model of the Josephson junction \label{II}}
\begin{figure}[t]
\includegraphics[width=\columnwidth]{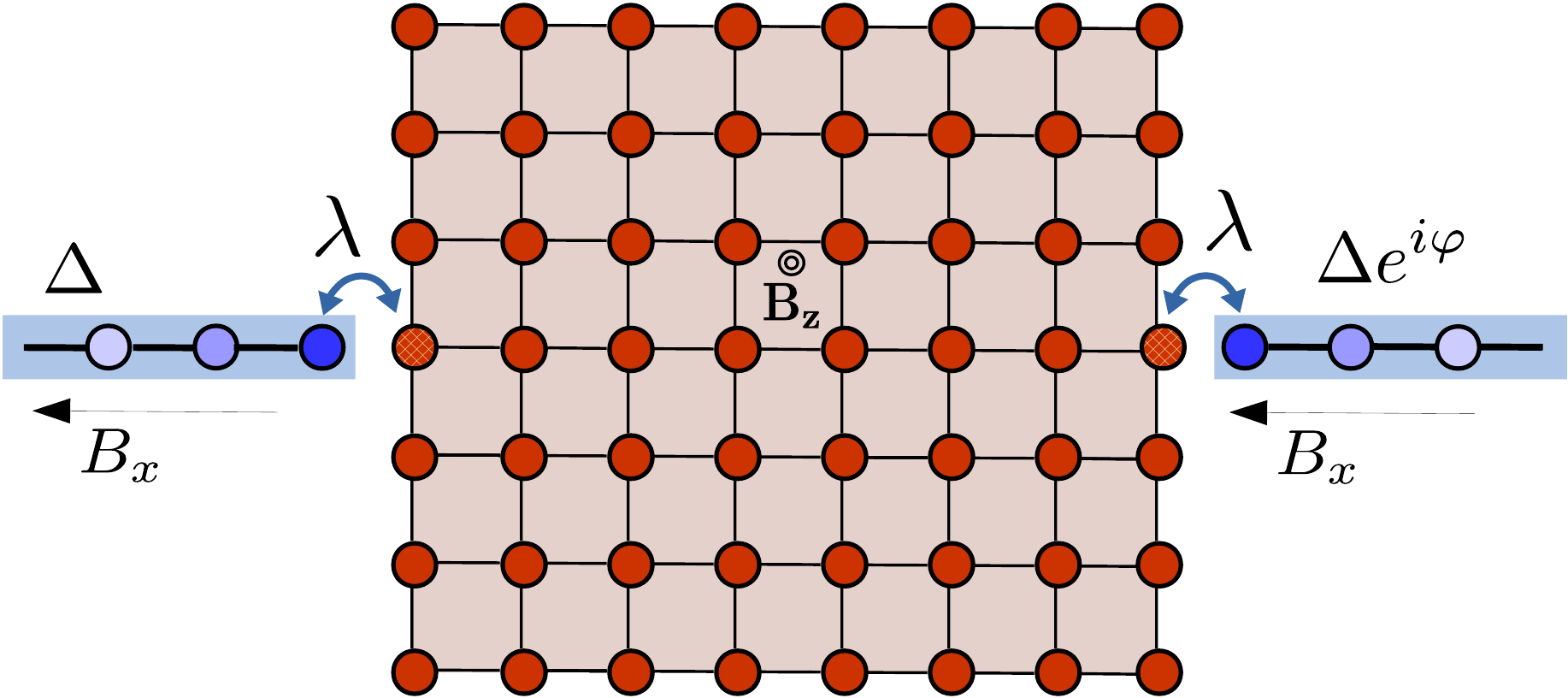}
\caption{Tight-binding scheme of the Josephson junction. Two nanowires with Rashba spin-orbit coupling proximitized with a BCS superconductor are subject to a Zeeman field in the $\vs{x}$ direction. The wires are coupled to a square lattice in the quantum Hall regime with a hopping amplitude $\lambda$.}
\label{fig1}
\end{figure}

We consider the system schematically shown in Fig.~\ref{fig1}. The quantum Hall (QH) central region is modeled with a square lattice threaded by a net geometrical flux $\Phi_g=B_zA_g$, where $B_z$ is the component of the applied magnetic field perpendicular to the lattice, and $A_g$ is the geometric area of the latter. We  use the Bogoliubov-de Gennes basis and describe the fields at each site $\bm{r}$ as $\hat{\Psi}_{\bm{r}} = (c_{\bm{r}\uparrow}^{}, c_{\bm{r}\downarrow}^{}, c^{\dagger}_{\bm{r}\downarrow}, -c^{\dagger}_{\bm{r}\uparrow})^{\mathrm{T}}$, where $c^{\dagger}_{\bm{r}\sigma}$ creates an electron with spin $\sigma$ at site $\bm{r}=x\,\vs{x}+y\,\vs{y}$ of the QH region. Taking the lattice spacing to be $a_0$, the Hamiltonian can be written as
\begin{eqnarray}
\notag
\hat{H}_{qh} &=& \frac{1}{2}\sum_{\bm{r}}\Big[\hat{\Psi}^{\dagger}_{\bm{r}}\mathcal{H}_{0} \hat{\Psi}^{}_{\bm{r}}+\hat{\Psi}^{\dagger}_{\bm{r}}V_{\bm{r},\bm{r}+a_0\hat{\bm{x}}}^{}\hat{\Psi}^{}_{\bm{r}+ a_0 \hat{\bm{x}}}\\
&+&\hat{\Psi}^{\dagger}_{\bm{r}}V_{\bm{r},\bm{r}+a_0\hat{\bm{y}}}^{}\hat{\Psi}^{}_{\bm{r}+ a_0 \hat{\bm{y}}} + h.c\Big]\,,
\end{eqnarray}
where
\begin{eqnarray}
\notag
\mathcal{H}_0 &=& (4t_{qh}-\mu - V_g)\,\tau_z\otimes\sigma_0\,,\\
\notag
V_{\bm{r},\bm{r}+a_0\hat{\bm{y}}} &=&-t_{qh}\,\tau_z\otimes\sigma_0\,,\\
V_{\bm{r},\bm{r}+a_0\hat{\bm{x}}} &=&-t_{qh}\,\tau_z\otimes\sigma_0\, e^{-i\frac{2\pi B_z y a_0}{\Phi_0}\tau_z\otimes\sigma_0}\,.
\end{eqnarray}
The hopping amplitude between neighboring sites  is given by $t_{qh}$, the chemical potential by $\mu$ and $V_g$ is a gate voltage that tunes the filling factor in the QH region. The Pauli matrices $\tau_{a}$ ($\sigma_{a}$) and the identity $\tau_0$ ($\sigma_0$) act in particle-hole (spin) space. $B_z$ has been included via the Peierls substitution, with the vector potential $\bm{A} = - B_z y\,\hat{\bm{x}}$  in the Landau gauge and the $y$ coordinate taken to be zero exactly at the middle of the sample (where the superconducting leads are attached). The Zeeman term in the QH region is assumed to be negligible.

The superconducting leads are modeled  as nanowires with Rashba spin-orbit coupling subject to an in plane Zeeman field $B_x$ and in proximity with a BCS superconductor of gap $\Delta$. As originally discussed in  Refs.~\cite{Lutchyn2010} and \cite{Oreg2010}, for $B_x$ larger than the critical field $B_c = \sqrt{\Delta^2 + \mu^2}$, with $\mu$ the chemical potential of the wires, topologically protected zero-energy Majorana modes arise at the ends of each lead. The model is well described by the following $N$-site one-dimensional lattice Hamiltonian:
\begin{equation}
\hat{H}_{\nu}\!=\!\frac{1}{2}\!\sum_{j=0}^{N-1}\!\hat{\chi}^{\nu\dagger}_{j}\mathcal{H}_{\nu}^{}\hat{\chi}^{\nu}_{j} + \!\frac{1}{2}\!\sum_{j=0}^{N-2}\!\Big[\hat{\chi}^{\nu\dagger}_{j}T_{\nu}^{}\hat{\chi}^{\nu}_{j+1} + \hat{\chi}^{\nu\dagger}_{j+1}T_{\nu}^{\dagger}\hat{\chi}^{\nu}_{j}\Big]\,,
\label{eqH}
\end{equation}
with
\begin{eqnarray}
\notag
\mathcal{H}_{\nu}&=& (2t_{sc} - \mu)\,\tau_z\otimes\sigma_0 - B_x\,\tau_0\otimes\sigma_x + \Delta\,\tau_x\otimes\sigma_0\, ,\\
T_{\nu} &=& -t_{sc}\,\tau_z\otimes\sigma_0 + i\alpha\,\tau_z\otimes\sigma_z\, .
\end{eqnarray}
Here $\nu = L, R$ refers to the left and right leads, and the four component spinor at site $j$ is merely ${\hat{\chi}^{\nu}_{j}}{} = (c_{j\uparrow}^{\nu},c_{j\downarrow}^{\nu},c_{j\downarrow}^{\nu\dagger},-c_{j\uparrow}^{\nu\dagger})^\mathrm{T}$. The hopping matrix element of the wires is given by $t_{sc}$, and $\alpha$ represents the spin-orbit coupling. For the purposes of this work, the number of sites $N$ is taken sufficiently large so the Majorana modes at opposite edges of each wire have negligible overlap. We can label the fields that will ultimately be coupled to the Hall bar by $\hat{\chi}^{L}_{N-1} \equiv \hat{\chi}_{L}$ and $\hat{\chi}^{R}_{0} \equiv \hat{\chi}_{R}$. The tunneling Hamiltonian between the leads and the central region is then given by
\begin{equation}
\hat{H}_T = \frac{1}{2}[\hat{\chi}^{\dagger}_L V_{L,\bm{r}_L}^{}\hat{\Psi}_{\bm{r}_{L}}+ \hat{\chi}^{\dagger}_R V_{R,\bm{r}_R}^{}\hat{\Psi}_{\bm{r}_R} + h.c.]\,,
\end{equation}
where we have incorporated the junction's phase difference $\varphi$ in the hopping to the right superconductor, and we defined $V_{L,\bm{r}_L}=-\lambda \,\tau_z\otimes\sigma_0$ and $V_{R,\bm{r}_R} = -\lambda\, e^{i\frac{\varphi}{2}\tau_z\otimes\sigma_0}\tau_z\otimes\sigma_0$. Here the coordinates $\bm{r}_L$ and $\bm{r}_R$ correspond to the sites at the edge of the Hall sample that are coupled to the left and right leads, respectively.
In our numerical simulations, we choose parameters such that the hoppings $t_{sc} = t_{qh} = \lambda = 1$, $\mu=0$, $\Delta=0.3$ and $\alpha=0.1$. A small square lattice of $N_y=41$ sites wide and $N_x=65$ sites long is used, so that the total geometrical area of the sample is $A_g=(N_x-1)(N_y-1)a_0^2$.
\subsection{Supercurrent and Fraunhofer patterns}
The current-phase relation of a Josephson junction is intrinsically endowed with valuable information on the mechanisms that build up the supercurrent. During the last few years, it has been particularly studied to disclose the presence of Majoranas in junctions with topological superconductors~\citep{Oreg2010,Lutchyn2010, Wiedenmann2016,Laroche2019, Aguado2017}. The critical current, defined as the maximum current in the current-phase relation, also provides relevant details on the physical processes that occur in the junction. In fact, its behavior when threading the region between superconductors with an out-of-plane magnetic field $B_z$ has been widely used as a tool to understand the nature of the supercurrent flow. When varying $B_z$, the magnetic flux threading the sample imposes a winding of the superconducting phase that results in modulations of the critical current, known as the Fraunhofer interference patterns. 

For the simplest case of a rectangular junction of area $A_g$, with a spatially homogeneous supercurrent density~\citep{Dynes1971},  the critical current is theoretically predicted to be given by
 $J_c (B_z)=J_c (0) |\sin(2 \pi \Phi_g/\Phi_0)/( 2 \pi \Phi_g/\Phi_0)|$ \cite{Tinkham1996} . Deviations from this result are known to occur in devices with inhomogeneities, such as non-uniform magnetic susceptibilites~\citep{Brcsk2019}, or when the magnetic field amplitude is enough to lead the system to a semiclassical regime, where electrons and holes deflect their paths in cyclotron orbits extending across the junction~\citep{BenShalom2015}. Within this scenario, irregular critical current profiles bearing aperiodic modulations or significantly enhanced or suppressed lobes are expected to occur. Under those circumstances, the transport properties are strongly dependent on the junction's geometry. Conversely, when the supercurrent is carried by edge states, a more regular and periodic pattern is expected. This has proven to be the case in quantum Hall~\citep{vanOstaay2011} and topological-insulator-based junctions~\citep{Pribiag2015,Baxevanis2015}, or even when trivial edge channels bridge the superconductors~\citep{deVries2018}. 

In what follows, we will then focus on the extreme quantum limit of our QH junction, where only the first Landau level is occupied.  Within the range of parameters we work with, a typical flux per plaquette of the order of $B_z a_0^2/\Phi_0 \simeq 0.08$ and a gate voltage $V_g=1$ are enough to satisfy this last condition. The magnetic length is such that $l_B \simeq 1.4\,a_0$ so that the edge states are sufficiently localized around the perimeter of the sample.

The zero temperature supercurrent flowing from the left superconductor to the Hall bar in equilibrium is obtained as
\begin{equation}
\langle\hat{J}_L\rangle = -\frac{e}{h}\,\mathrm{Re}\int d\omega\,\mathrm{Tr}\left[\tau_z\otimes\sigma_0\, V_{L,\bm{r}_L}^{}G^{<}_{\bm{r}_L,L}(\omega)\right]\,,
\label{current}
\end{equation}
where $G^{<}_{\bm{r}_L,L}(\omega)$ is the minor Green's function between the left coupled site of the Hall bar and the corresponding lead. Its elements in the Bogoliubov-de Gennes basis are defined as  $[G^{<}_{\bm{r}_L,L}(\omega)]_{\alpha\beta} = i\int dt e^{i\omega t }\langle\hat{\chi}^{\dagger}_{L\beta}(0)\hat{\Psi}_{\bm{r}_L\alpha}^{}(t)\rangle$ \citep{Jauho1996}, and, in equilibrium, it satisfies a simple relation with  the retarded ($r$) and advanced ($a$) Green's functions
\begin{equation}
G^{<}_{\bm{r}_L,L}(\omega) = f(\omega)\left[G^{a}_{\bm{r}_L,L}(\omega)-G^{r}_{\bm{r}_L,L}(\omega)  \right],
\label{Gminor}
\end{equation}
with $f(\omega)$ the Fermi-Dirac distribution, which is taken here to be a Heaviside function at zero temperature,  $f(\omega)=\Theta(-\omega)$. One should bear in mind that, within this formalism, the supercurrent is always $2\pi$-periodic on account of this thermodynamic average without parity conserving constraints~\footnote{This has been proven to be the most likely scenario, mainly because of the ubiquitous presence of quasiparticle poisoning in experimental devices.}. 

\begin{figure}[t]
\includegraphics[width=0.95\columnwidth]{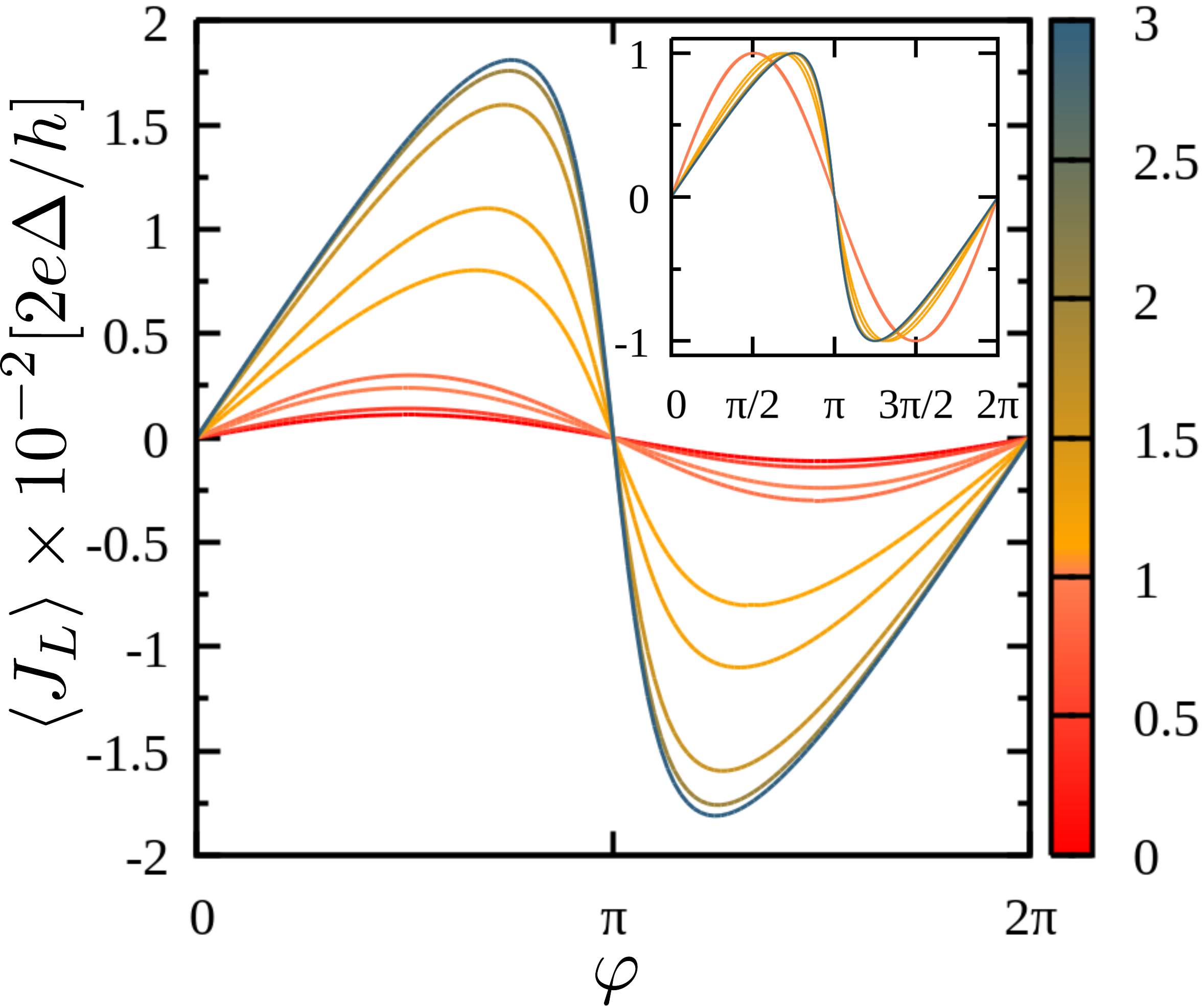}
\caption{Current-phase relations for different magnetic fields along the nanowires. The total geometrical flux threading the sample is given by $\Phi_g = 0.08 \Phi_0 \frac{A_g}{a_0^2}$. The color scale indicates the magnitude of the Zeeman field along the wires normalized to the critical field $B_x/B_c$. The inset shows the same curves normalized to their maximum value.}
\label{cpr}
\end{figure}

In Fig.~\ref{cpr} we show the current-phase relations in the quantum Hall regime calculated for different Zeeman fields ($B_x$) along the superconducting wires. The total geometrical flux in the Hall sample is chosen to be $\Phi_g=0.08\Phi_0\frac{A_g}{a_0^2}$. Notice that our choice for the vector potential gauge  and the symmetrical positioning of the superconducting leads guarantees the absence of a supercurrent at zero phase difference. An increment of the critical current in around an order of magnitude as the leads are driven throughout the topological phase transition is apparent from the figure. This phenomenon stems from an enhancement of the Andreev process when Majorana zero energy quasiparticles emerge at the end sites of each lead, as will be explained in Section \ref{III}. The changes in the current-phase relations profiles can be better visualized in the inset of Fig. \ref{cpr}, where each current has been normalized to its critical value. The maximum value of the curves shifts from being at $\varphi=\pi/2$ in the trivial phase to being closer to $\varphi=\pi$ in the topological phase. This effect is expected in the presence of Majoranas because the Andreev level spectrum becomes gapless. In particular, a topologically protected crossing between these bound states occurs when the phase difference between the superconducting nanowires is $\varphi=\pi$, which explains the aforementioned shift in the maximum critical current.
\begin{figure}[t]
\includegraphics[width=0.95\columnwidth]{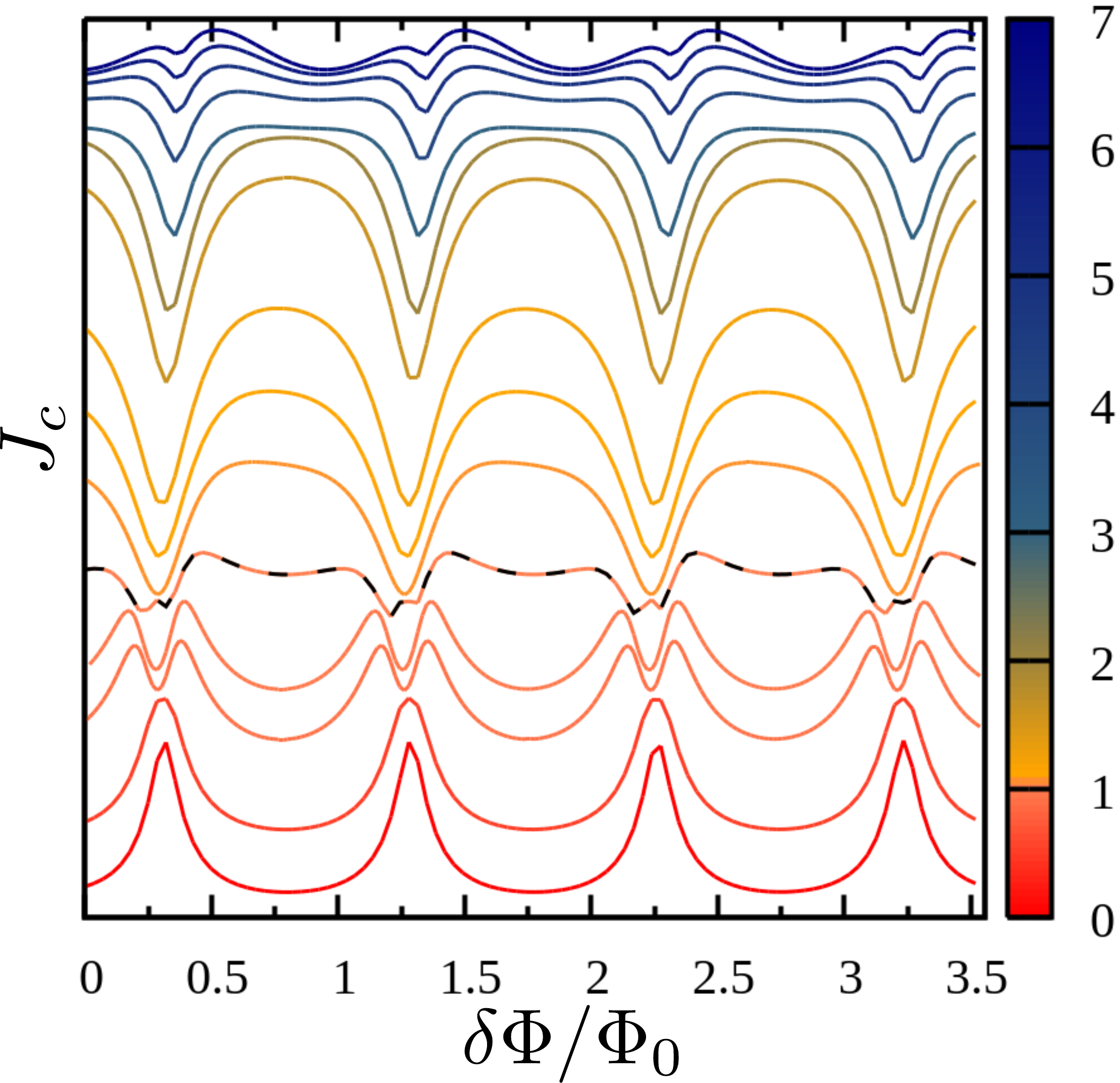}
\caption{Fraunhofer patterns: We show the critical current profiles as a function of the variations of total flux enclosed by the edge state $\delta\Phi$ relative to an initial flux $\Phi_g$ that sets the system safely into the quantum Hall regime. The color scale indicates the magnitude of the Zeeman field along the wires normalized to the critical field, $B_x/B_c$. Dashed lines indicate the Fraunhofer pattern at the topological phase transition $B_x= B_c$. The curves are shifted for clarity.}
\label{fraunhofer}
\end{figure}

Figure \ref{fraunhofer} shows the numerically obtained Fraunhofer patterns, calculated as
\begin{equation}
J_c(\delta B_z) = \max\limits_{\varphi}| \langle J_L(\varphi,\Phi_g+\delta B_z A_g) \rangle|\,,
\end{equation}
where we change the magnetic field threading the central region in $\delta B_z$. Each curve has a different magnitude of the Zeeman field $B_x$, represented by a color scale normalized to the critical field $B_c$. 
To properly compare the flux variation  $\delta\Phi$ with the flux quantum, the former is calculated as  $\delta\Phi=\delta B_z A_\mathrm{ph}$, where $A_\mathrm{ph}=[(N_x-1)a_0-2l_B][(N_y-1)a_0-2l_B]$ is the physical area enclosed by the edge state, which has  a typical size of $l_B \simeq 1.4 a_0$.

At $B_x=0$ we obtain the already known characteristic Fraunhofer profile of a supercurrent carried by a chiral edge state \cite{Ma1993,vanOstaay2011} with a periodicity given by the normal flux quantum $\Phi_0$. The presence of peaks or resonances can be traced back to the level discretization of the chiral edge state due to its confinement along the perimeter of the Hall bar. Each time one of these discrete levels becomes resonant with the Fermi level\textemdash a condition which is naturally periodic with $\Phi_0$\textemdash the supercurrent becomes larger in magnitude. As $B_x$ gets closer to the critical value, these resonances are spin-split: since the effective superconducting gap is reduced, the bound Andreev levels penetrate deeper into the leads and hence the effect of the Zeeman coupling becomes stronger.

\begin{figure}[t]
\includegraphics[width=0.95\columnwidth]{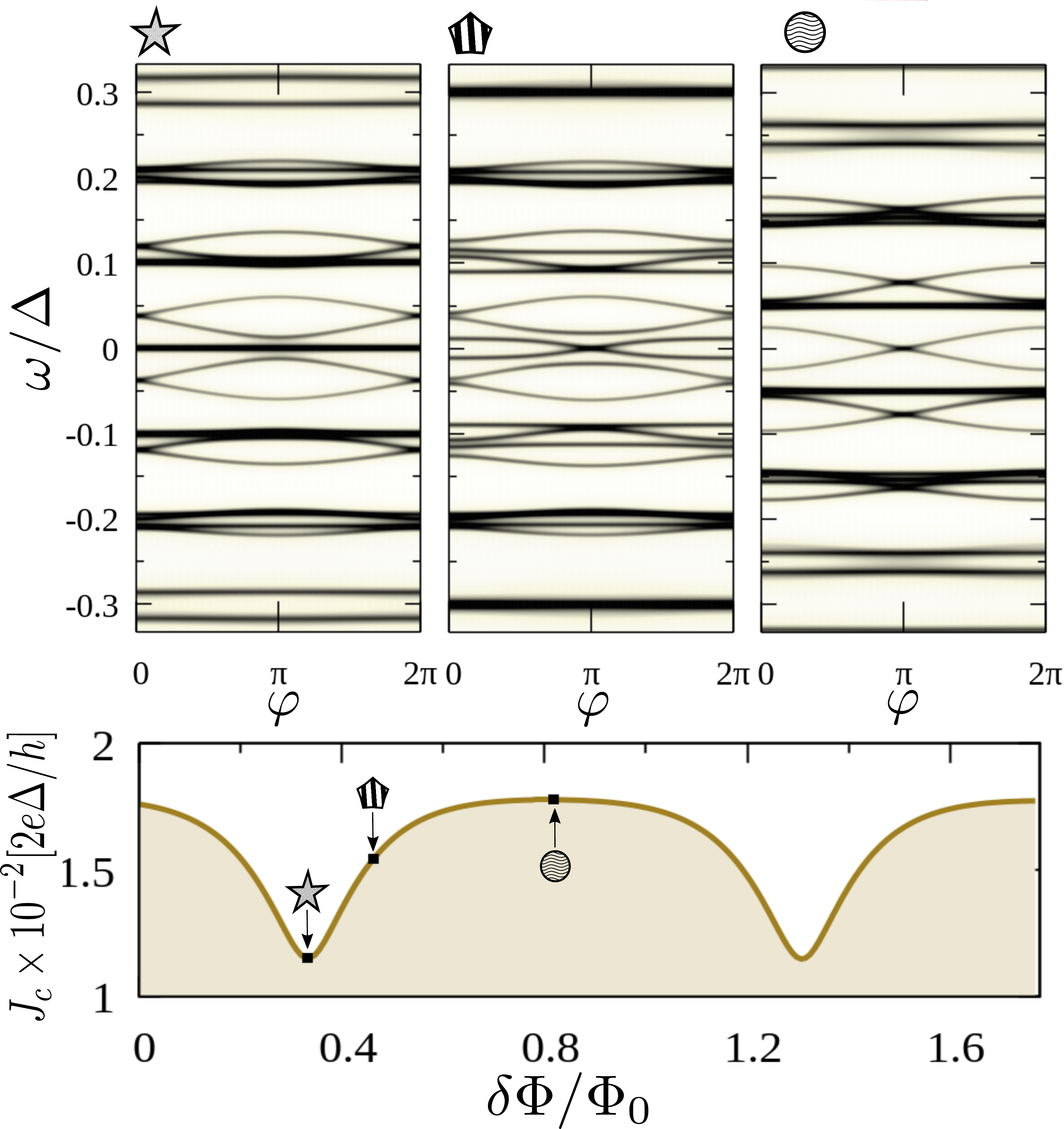}
\caption{The upper panels show the spectral density $\mathcal{A}_L(\omega,\varphi)$ in the quantum Hall regime. The color scale goes from white (zero) to black (higher value) in arbitrary units. Each figure is calculated for a different magnetic flux threading the central region, indicated by a corresponding symbol in the  Fraunhofer pattern shown in the lower panel. The nanowires are in the topological regime with a magnetic field $B_x = 2 B_c$. }
\label{absBx2-0}
\end{figure}

\begin{figure}[t]
\includegraphics[width=0.95\columnwidth]{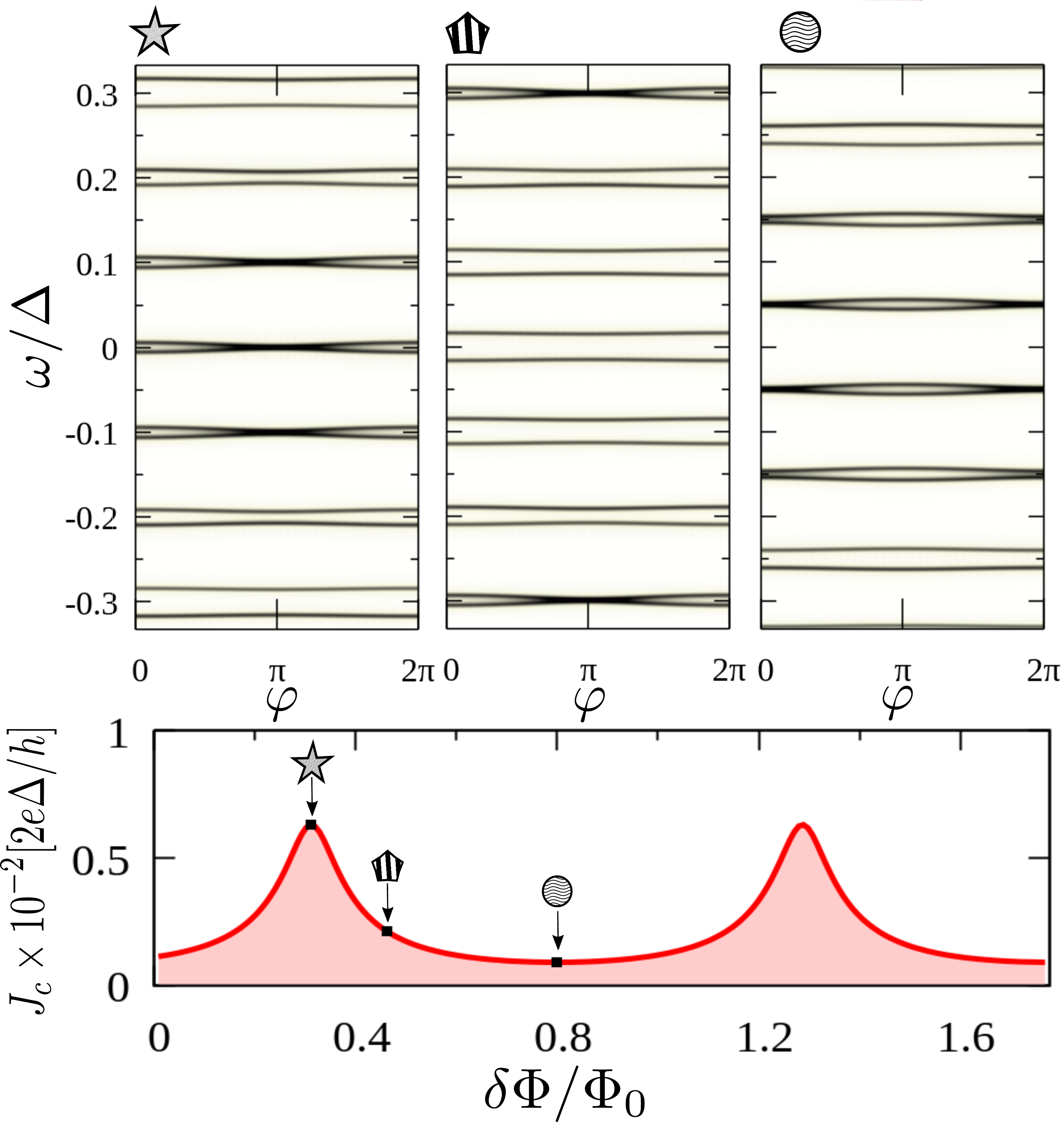}
\caption{Same as the previous figure but with a magnetic field $B_x = 0$ so that  the nanowires are in the trivial regime. }
\label{absBx0-0}
\end{figure}
Quite remarkably, the Fraunhofer patterns change drastically in the topological phase, i.e., for fields $B_x > B_c$. Even though the periodicity in the oscillations remains the same, the resonances have now become dips in the critical current profiles. These dips have an additional field-dependent magnitude: they tend to smoothly disappear as $B_x$ is further increased. As we shall explain in section \ref{III}, this behavior can be understood as a clear signature of the presence of Majorana fermions at the ends of each lead. Interestingly, the spin polarization of these topologically protected quasi-particles is found to be responsible for the above mentioned magnetic field dependence   of the Fraunhofer profiles.  

We also note in passing the absence of nodes (zeros) in the critical current patterns both in the topological and the trivial phase, as opposed to the Fraunhofer oscillations in a conventional Josephson junction~\citep{Tinkham1996}. This effect has also been pointed out to occur in a quantum spin Hall based junction hosting one-dimensional topological superconductivity~\cite{Lee2014}.

\subsection{Andreev bound states}

To understand the transport properties of the junction, it is instructive to take a closer look at the Andreev bound states, which generally carry most of the supercurrent between superconductors. In order to do so, we calculate the spectral density at the left edge of the central region as
\begin{equation}
\mathcal{A}_L(\omega,\varphi) = -\frac{1}{\pi}\text{Im}\sum_{\bm{r}\,\epsilon\,L}\text{Tr}\Big[G^{r}_{\bm{r}\bm{r}}(\omega,\varphi)\Big],
\end{equation}
where $G^{r}_{\bm{r}\bm{r}}$ is the retarded Green's function of the field $\hat{\Psi}_{\bm{r}}$. This magnitude faithfully reveals the Andreev bound-states dispersion relation as a function of the phase difference between the superconducting leads.
In Figs. \ref{absBx2-0} and \ref{absBx0-0} we show the behavior of $\mathcal{A}_L(\omega,\varphi)$ when the nanowires are in the the topological regime ($B_x=2 B_c$)  and in the trivial regime ($B_x = 0$), respectively. The parameters of the quantum Hall region are the ones used for the transport simulations in the previous section. We have chosen three significant fluxes in the Fraunhofer patterns, shown in the lower panels, to calculate the corresponding subgap spectral densities. 

Andreev bound states in this junction arise near the energies where discrete levels are formed due to the confinement of the chiral edge state in the perimeter of the isolated Hall bar, bearing a resemblance to the ones obtained in the case of a one-dimensional channel between superconductors. In fact, they come in sets that are determined by the level spacing $\delta\varepsilon = 2\pi\hbar v_d/P$, with $P$ the perimeter of the square lattice and $v_d$ the drift velocity of the edge state. For our chosen parameters, $\delta\varepsilon\simeq0.1\Delta$.

Some fingerprints in the spectral densities are clearly correlated with the magnetic flux dependence of the Fraunhofer patterns. In the topological case (Fig. \ref{absBx2-0}), when a dip occurs in the critical current profile, a series of low energy levels become non-dispersive and degenerate in pairs. In particular, two levels stay pinned at the Fermi level. As we shall explain in section \ref{III}, this effect originates when four degenerate levels (taking into account the electronic and hole sectors as well as their spins) are coupled to the zero energy Majorana modes. In this situation, it is always possible to find two linear combinations of these states that effectively decouple from the leads.  When the flux is detuned from this particular point, the levels become dispersive, naturally translating into a larger critical current. At phase difference $\varphi=\pi$ a topologically protected crossing occurs between these levels, in a similar fashion to the case of a tunnel junction between two Majorana fermions. The supercurrent becomes maximum when the flux is chosen to be in between two dips.  

In the trivial case  at $B_x=0$ (Fig. \ref{absBx0-0}) all Andreev levels are spin-degenerate. A resonance takes place in the critical current profile when the electronic and hole states become degenerate at the Fermi level. The superconducting correlations couple these levels, so they become dispersive as a function of the phase difference and eventually cross at $\varphi=\pi$, where the current becomes maximum. As the flux gets detuned from this value, this crossing becomes an anti-crossing and the Andreev bound states tend to be less dispersive, causing the value of the current to diminish. 
\section{Low energy spinful model \label{III}}
\begin{figure}[t]
\includegraphics[width=\columnwidth]{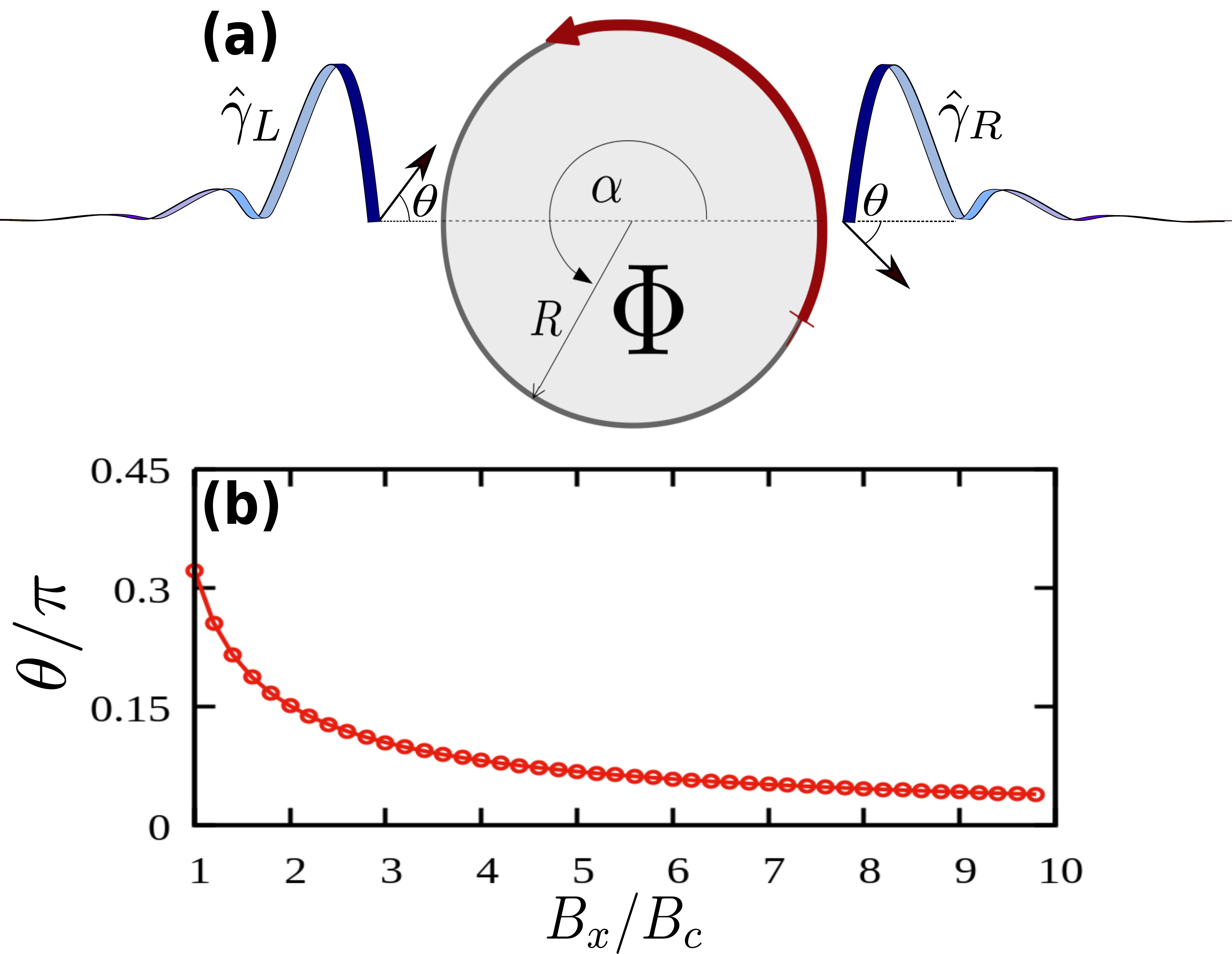}
\caption{(a) Low energy model of the Josephson junction: Two Majorana end modes are coupled with a chiral field with velocity $v_d$ that moves on a ring of radius $R$ threaded by an external flux $\Phi$. The arrows represent the spin of the quasiparticles at the end sites of each wire. The canting angle $\theta$ is defined in the main text. (b) Calculated spin canting angle as a function of the Zeeman field along the wire. We show results within the topological regime $B_x > B_c$. The parameters of the superconducting nanowire are specified in the main text.}
\label{modelfig}
\end{figure}

In this section we introduce a low energy spinful model, schematically depicted in Fig.~\ref{modelfig}, to understand the results in the topological regime ($B_x>B_c$). Two Majorana fermions, represented by the operators $\hat{\gamma}_L$ and $\hat{\gamma}_R$, are coupled with a hopping amplitude $\bar{\lambda}$ to a chiral one-dimensional channel with drift velocity $v_d$. The Hamiltonian describing the chiral field is described as
\begin{equation}
\hat{H}_{\text{ch}} = \hbar v_d \sum_{\sigma}\int_{0}^{2\pi} d\alpha\,\hat{\psi}^{\dagger}_{\sigma}(\alpha)(-i\partial_{\alpha}-\Phi/\Phi_0)\hat{\psi}_{\sigma}(\alpha)\,,
\label{eqchiral}
\end{equation}
where we have used angular coordinates to write the vector potential along the radius $R$ of the ring as $\bm{A} = \frac{B_z R}{2} \vs{\alpha}$. The net magnetic flux through the ring is $\Phi = B_z\pi R^2$. The chiral fields are normalized when integrated along the perimeter of the ring.
The tunneling Hamiltonian between the Majorana fermions and the one dimensional channel can be written as
\begin{eqnarray}
\notag
\hat{H}_T &=& \bar{\lambda} \int \, d\alpha\,\delta(\alpha)\, \hat{\gamma}_R \,\Big(e^{i\varphi/2}\hat{\psi}_1(\alpha)-e^{-i\varphi/2}\hat{\psi}^{\dagger}_{1}(\alpha)\Big)\\
&&+\bar{\lambda} \int\, d\alpha\,\delta(\alpha-\pi)\, \hat{\gamma}_L\,
\Big(\hat{\widetilde{\psi}}_1(\alpha) - \hat{\widetilde{\psi}}_1^{\dagger}(\alpha)\Big)\,,
\label{tunnH}
\end{eqnarray}
where the phase difference $\varphi$ between the superconducting leads has been taken fully into account only on the hopping to the right Majorana fermion and we have defined the fields
\begin{eqnarray}
\notag
\hat{\psi}_1(\alpha) &=& \cos(\theta/2)\,\hat{\psi}_{\uparrow}(\alpha)-i\sin(\theta/2)\,\hat{\psi}_{\downarrow}(\alpha)\\
\notag
\hat{\psi}_2(\alpha) &=& -i\sin(\theta/2)\,\hat{\psi}_{\uparrow}(\alpha)+\cos(\theta/2)\,\hat{\psi}_{\downarrow}(\alpha)\\
\notag
\hat{\widetilde{\psi}}_1(\alpha) &=& i\cos(\theta/2)\,\hat{\psi}_{\uparrow}(\alpha)-\sin(\theta/2)\,\hat{\psi}_{\downarrow}(\alpha)\\
\hat{\widetilde{\psi}}_2(\alpha) &=& \sin(\theta/2)\,\hat{\psi}_{\uparrow}(\alpha)-i\cos(\theta/2)\,\hat{\psi}_{\downarrow}(\alpha)\,.
\end{eqnarray}

Here, the spin quantization axis ($\uparrow,\downarrow$) has been defined parallel to that of the magnetic field along the wires ($\hat{x}$). We have chosen this particular form of the coupling so as to preserve the spin degree of freedom in the tunneling Hamiltonian. Since the spin-orbit effective field of the original wires is in the $\hat{z}$ direction [see Eq. (\ref{eqH})], the spin polarization of the left and right Majorana fermions lays on the $x-y$ plane~\citep{Sticlet2012}. In particular, both quasiparticles have the same spin projection along the  direction of the Zeeman field but bare a different sign along the $\hat{y}$ direction. This behavior is captured by the canting angle $\theta$.  The $\hat{\psi}_2(\alpha)$ and $\hat{\widetilde{\psi}}_2(\alpha)$ fields do not appear in the tunneling Hamiltonian since their spins are anti-parallel to the right and left Majorana fermions, respectively. 
 
We can gain a better insight by computing $\theta$ for the parameters used in our tight-binding numerical simulations. Specifically,
\begin{equation}
\tan(\theta) = \lim_{\omega\to 0}\frac{\langle\hat{S}_y(\omega)\rangle}{\langle\hat{S}_x(\omega)\rangle}\,,
\end{equation}
where 
\begin{equation}
\langle\hat{S}_j(\omega)\rangle = -\frac{\hbar}{2\pi}\text{Im}\text{Tr}\Big[\frac{(\tau_0+\tau_z)}{2}\sigma_j G^{r}_{\hat{\chi}_{L}}\Big],
\end{equation}
with $G^{r}_{\hat{\chi}_{L}}$ the retarded Green's function of the end site of the left lead. We plot this angle in Fig.~\ref{modelfig}(b) as a function of the Zeeman field $B_x$. For high fields, the spins tend to be completely aligned along the $\hat{x}$ direction, ultimately arriving to the well known Kitaev ``spinless" limit~\citep{Kitaev2001}. 

\subsection{Andreev bound states}
Our first purpose is to find the bound states of the model defined by Eqs. (\ref{eqchiral}) and (\ref{tunnH}). A natural way of integrating out the Majorana fermions from the tunneling Hamiltonian is to solve the scattering problem of the chiral fermions at each terminal. The incoming electronic/hole modes with energy $\omega$ at the right lead ($\alpha = 0^{-}$) are related to the outgoing ones ($\alpha = 0^{+}$) through the transfer matrix $M_0(\omega,\varphi)$
\begin{equation}
\hat{\Psi}(0^{+}) = M_0(\omega,\varphi)\hat{\Psi}(0^{-})\,.
\label{m0}
\end{equation} 
with $\hat{\Psi}(\alpha) = (\hat{\psi}_1(\alpha), \hat{\psi}_1^{\dagger}(\alpha), \hat{\psi}_2(\alpha), \hat{\psi}_2^{\dagger}(\alpha))^{T}$  \footnote{Note that here we exchanged the particle-hole and spin subspaces as compared with the notation used in the tight-binding model}, and 
\begin{equation}
M_0(\omega,\varphi) = \begin{pmatrix}
\frac{\hbar R v_d \omega}{2 i  \bar{\lambda}^2 + \hbar R v_d \omega} & \frac{2 i \bar{\lambda}^2 e^{-i\varphi}}{2 i \bar{\lambda}^2 + \hbar R v_d \omega} & 0 & 0\\
\frac{2 i  \bar{\lambda}^2 e^{i\varphi}}{2 i  \bar{\lambda}^2 + \hbar R v_d \omega} & \frac{\hbar R v_d \omega}{2 i  \bar{\lambda}^2 + \hbar R v_d \omega} & 0 & 0\\
0 & 0 & 1 & 0\\
0 & 0 & 0 & 1
\end{pmatrix}\,.
\end{equation}
The peculiarity of this type of scattering is that if a zero energy electron (hole) with a spin parallel to the one of $\hat{\psi}_1$ scatters with the Majorana mode, a perfect Andreev reflection takes place and a hole (electron) with the same spin as the incoming particle goes through. This phenomenon is known as the selective equal spin Andreev reflection~\citep{He2014}.

The transfer matrix in the left lead $M_{\pi}(\omega)$ can be written as a rotation of the one obtained for the right lead,
\begin{equation}
M_{\pi}(\omega) = C^{\dagger}(\theta)M_0(\omega,0) C(\theta)\,,
\label{mpi}
\end{equation}
with $C(\theta)$ defined by
\begin{equation}
C(\theta) = \begin{pmatrix}
i \cos(\theta) & 0 & -\sin(\theta) & 0\\
0 & -i\cos(\theta) & 0 & -\sin(\theta)\\
\sin(\theta) & 0 & -i\cos(\theta) & 0\\
0 & \sin(\theta) & 0 & i\cos(\theta)
\end{pmatrix}\,.
\end{equation}
The matrix $C(\theta)$ is such that $\hat{\widetilde{\Psi}}(\alpha) = C(\theta)\hat{\Psi}(\alpha)$, with $\hat{\widetilde{\Psi}}(\alpha) = (\hat{\widetilde{\psi}}_1(\alpha), \hat{\widetilde{\psi}}_1^{\dagger}(\alpha), \hat{\widetilde{\psi}}_2(\alpha), \hat{\widetilde{\psi}}_2^{\dagger}(\alpha))^{T}$. A straightforward  piece-wise integration of the Schr$\ddot{\text{o}}$dinger equation defined by Eq. (\ref{eqchiral}), with the boundary conditions (\ref{m0}) and (\ref{mpi}) shows that an eigenstate at $\alpha=0^{+}$ with energy $\omega$ must then satisfy 
\begin{equation}
\hat{\Psi}(0^{+}) = \Pi(\omega)\hat{\Psi}(0^{+}),
\label{eigenstates}
\end{equation}
with
\begin{equation}
\Pi(\omega)=e^{i2\pi \tilde{\omega}} M_0(\omega,\varphi) e^{i\pi\tilde{\Phi}\,\sigma_0\otimes\tau_z} M_{\pi}(\omega)e^{i\pi\tilde{\Phi}\,\sigma_0\otimes\tau_z}.
\end{equation}
Here $\tilde{\omega}=\omega/\delta\varepsilon$, $\delta\varepsilon=\hbar v_d/R$ is the level spacing of the chiral modes in the ring and $\tilde{\Phi}=\Phi/\Phi_0$. The eigenenergies of the system are then given by
the equation
\begin{equation}
\mathrm{det}[\mathbb{1}-\Pi(\omega)]=0\,.
\label{det}
\end{equation}
When $\bar{\lambda}=0$ it is trivial to obtain the electronic ($-$) and hole ($+$) spin-degenerate solutions of the uncoupled ring $\omega_n^{\pm} = \delta\varepsilon\, (n \pm \Phi/\Phi_0)$, with $n\in \mathbb{Z}$. Note that with our choice of zero chemical potential in Eq. (\ref{eqchiral}) there is always an electronic and a hole mode at the Fermi energy whenever there is an integer number of flux quanta threading the system.

\begin{figure}[t]
\includegraphics[width=\columnwidth]{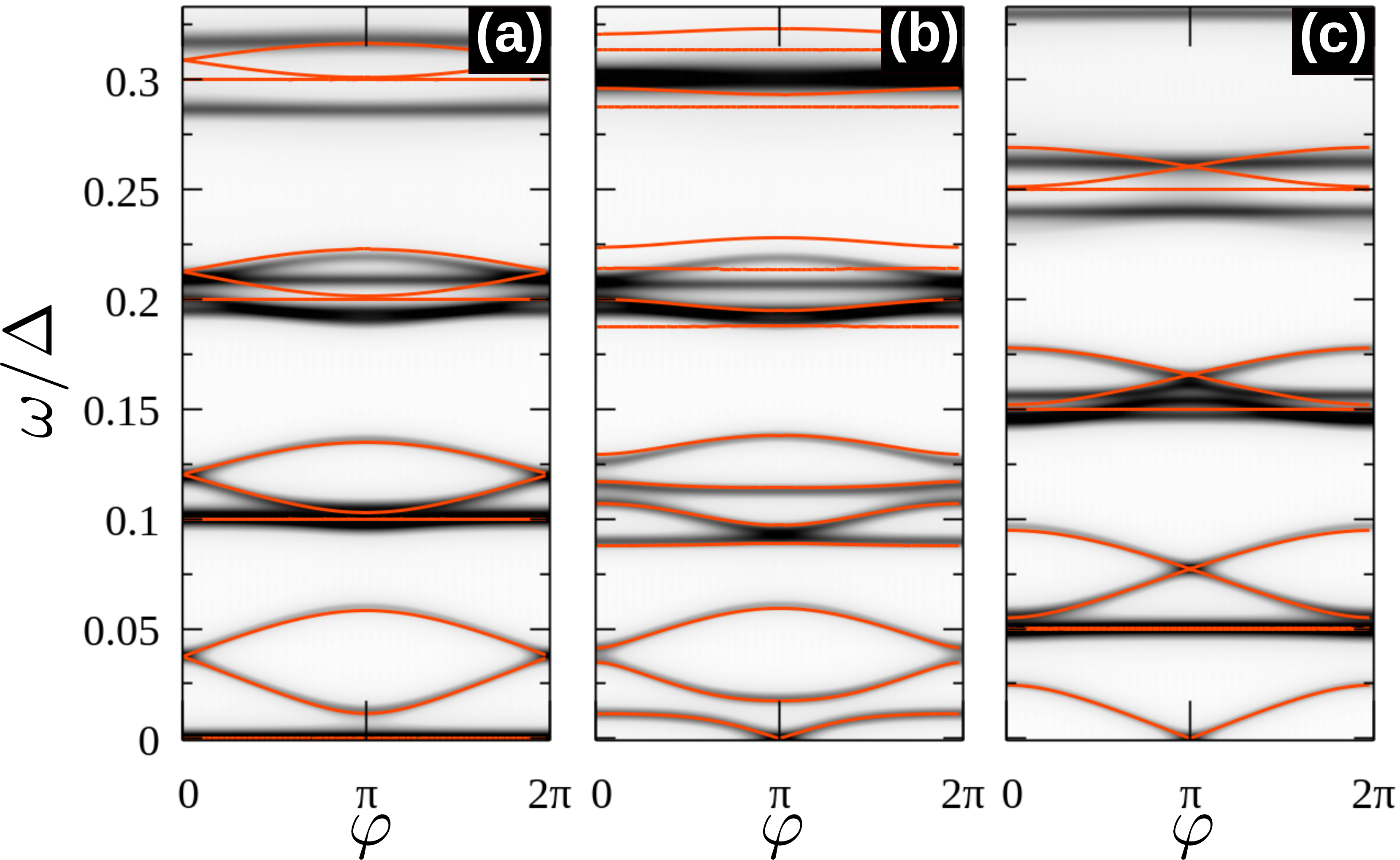}
\caption{Solid red lines are the analytical solutions obtained from Eq. (\ref{det}) as a function of the phase difference $\varphi$ between superconductors. In all figures $R=1$, $\bar{\lambda}=0.02$, $\theta=0.15\pi$ and $\delta\varepsilon = 0.03$. In (a) there is an integer number of flux quantum threading the ring, in (b) $\Phi = 0.13 \Phi_0$ and in (c) $\Phi=0.5\Phi_0$. The color maps are the spectral densities $\mathcal{A}_L(\omega,\varphi)$ for $B_x = 2 B_c$ and the same fluxes of Fig. \ref{absBx2-0}.}
\label{abs_analytical}
\end{figure}

Figure  \ref{abs_analytical} shows  the solutions of Eq. (\ref{det})  as a function of the phase difference $\varphi$ (solid (red) lines). Only positive eigenenergies are shown since the spectrum is electron-hole symmetric.  In all figures the radius of the ring is $R=1$, the hopping amplitude $\bar{\lambda}=0.02$, the level spacing $\delta\varepsilon = 0.03$ and the canting angle $\theta=0.15\pi$. These values where chosen so as to compare the results with the ones analyzed in Fig. \ref{absBx2-0} of Section \ref{II}, with a Zeeman field $B_x = 2 B_c$, while the canting angle has been extracted from Fig. \ref{modelfig}(b). In Fig. \ref{abs_analytical}(a) the flux is an integer number of $\Phi_0$, in (b) $\Phi=0.13\,\Phi_0$ and in (c) $\Phi=0.5\,\Phi_0$. Clearly, there is a good agreement between the theory and the tight-binding numerical simulations at low energies. At higher energies the model fails to describe the full spectral density because of two main reasons: (i) the assumption of an unrestricted linear spectrum for the edge state with a constant slope $v_d$ is not longer valid; (ii) the fact that the continuum spectrum has not been taken into account. Yet, as we shall show below, the low energy description is enough to qualitatively capture the main features of the complete transport simulations.

When the flux threading the ring is an integer number of flux quanta $\tilde{\Phi} = N$ the electronic and hole levels of the uncoupled system become degenerate at energies $n\delta\varepsilon$ since $\omega^{+}_{n-N} = \omega^{-}_{n+N}$. Taking into account the spin degree of freedom, this results in four degenerate states (for each $n$) coupled to two zero energy Majorana modes. This being the case, there are always two solutions that stay pinned at $n\delta\varepsilon$\footnote{This is a general result from linear algebra: an $n\times n$ hermitian matrix that contains a $m\times m$ degenerate subspace (with eigenvalue $\lambda$) has always at least $n-m$ eigenvectors inside the degenerate subspace that have eigenvalue $\lambda$.}, which are nothing but the series of flat bands in Fig. \ref{abs_analytical}(a). 
The two eigenstates at $\alpha=0$ corresponding to the $\omega=n\delta\varepsilon$ flat bands can be obtained from Eq.~(\ref{eigenstates}) as
\begin{eqnarray}
\notag
\Psi_{a}(0) &=& \mathcal{N}(\theta,\varphi)\left(i\frac{\tan(\theta)}{1 + e^{i\varphi}}, i\frac{\tan(\theta)}{1+e^{-i\varphi}}, 0, 1\right)\,,\\
\Psi_{b}(0) &=& \mathcal{N}(\theta,\varphi)\left(-i\frac{\tan(\theta)}{1 + e^{i\varphi}}, -i\frac{\tan(\theta)}{1 + e^{-i\varphi}}, 1, 0\right)\,,
\label{flat_states}
\end{eqnarray}
where the normalization factor is given by $\mathcal{N}(\theta,\varphi) = [1+\sec^2(\varphi/2)\tan^2(\theta)/2]^{-\frac{1}{2}}$. One can check that these states are eigenstates of both $M_0$ and $M_{\pi}$, and are therefore effectively decoupled from both Majorana fermions \footnote{Eq.~\eqref{flat_states} can also be found by inspection, looking for a linear combination of the $\psi_i^{}$ and $\psi_i^\dagger$  fields that do not couple to $\hat{\gamma}_L$ and $\hat{\gamma}_R$ (cf. Eq.~\eqref{tunnH}), taking into account that $\hat{\Psi}(\pi) = \hat{\Psi}(0)$ for $\tilde{\Phi}=N$.} and unable to carry supercurrent. 
This is consistent with the behavior of the dips in the Fraunhofer pattern discussed in Fig. \ref{absBx2-0}. As a matter of fact, the decoupled solutions are the ones that cease to contribute to the supercurrent when $\tilde{\Phi}$ is precisely tuned, resulting in a minimum in the critical current. Note that, as the canting angle $\theta$ decreases, the states in Eq.~\eqref{flat_states} tend to have a polarized spin parallel to that of the $\hat{\Psi}_2$ fields. When this happens, these solutions are always decoupled for any flux, so the dips disappear.

A similar scenario takes place when a half-integer number of flux quanta  $\tilde{\Phi} = (N+1/2)$ is threading the ring, since $\omega_{n-N}^{+} = \omega_{n+N+1}^{-}$. However, in this case, the  non-dispersive solutions will be at energies $\delta\varepsilon (n+\frac{1}{2})$, so that there is no flat band pinned at the Fermi energy---the closest to the Fermi level are at $\pm\delta\varepsilon/2$. Yet, as there is a topologically protected crossing at $\varphi=\pi$, there must be another pair of Andreev levels in between them, which  are maximally dispersive in that situation. This qualitative picture explains the maximum of $J_c$.  

\subsection{Josephson supercurrent}
In order to obtain the supercurrent, we first make a gauge transformation so that the phase difference between the topological superconducting  leads is taken into account by adding to Eq. (\ref{eqchiral}) the following contribution \cite{Alavirad2018}:
\begin{equation}
\hat{H}_{\varphi}= -\frac{\hbar v_d} {2}\!\int_{0}^{2\pi}\!d\alpha\,\hat{\Psi}^{\dagger}\Big(\frac{a(\alpha,\varphi)}{2}(\sigma_0 + \sigma_z)\otimes\tau_z\Big)\hat{\Psi}\,,
\end{equation}
with the vector potential $a(\alpha,\varphi)=\mathrm{sgn}(\alpha-\pi)\,\varphi/2\pi$. Notice that the phase dependent vector potential affects only the $\hat{\psi}_1$ fields, consistent with our initial choice.
The supercurrent is then given by $J_{\text{sc}}=\frac{2e}{\hbar}\langle\frac{\partial \hat{H}_{\varphi}}{\partial \varphi}\rangle$ and can be expressed as
\begin{eqnarray}
\notag
J_\text{sc} =&&\frac{e v_dk_B T}{4\pi}\sum_{m}\Big\{\int_{0}^{\pi}\!d\alpha\, \mathrm{Tr}\Big[(\sigma_0+\sigma_z)\otimes\tau_z \,\mathcal{G}(\alpha,\alpha,i\omega_m)\Big]\\
&&-\int_{\pi}^{2\pi}d\alpha\,\mathrm{Tr}\Big[(\sigma_0+\sigma_z)\otimes\tau_z \,\mathcal{G}(\alpha,\alpha,i\omega_m)\Big]\Big\},
\label{Jsc_G}
\end{eqnarray}
where $\mathcal{G}(\alpha,\alpha,i\omega_m)$ is the Matsubara Green's function of the chiral states, $\omega_m = (2m+1)\pi k_B T$ is the fermionic Matsubara frequency and $T$ the temperature. After the explicit evaluation of $\mathcal{G}(\alpha,\alpha,i\omega_m)$ (see Appendix \ref{Green}) we find 
\begin{widetext}
\begin{eqnarray}
\notag
J_{\text{sc}}(\varphi)&=&-\frac{ek_B T}{\hbar}\sum_m \Big\{2 \widetilde{\bar{\lambda}}^4 \cos^2(\theta)\sin(\varphi)\left[\cos(2 \pi\tilde{\Phi})-\cosh(2\pi\tilde{\omega}_m)\right]\Big\}\times
\Big\{\widetilde{\bar{\lambda}}^4 \left[\cos(2\theta)-2\cos^2(\theta) \cos(2 \pi\tilde{\Phi})\cos(\varphi)\right]
\\
\notag
&+&\left(\widetilde{\bar{\lambda}}^4+\pi ^2 \tilde{\omega}_m^2\right) \cosh (4\pi \tilde{\omega}_m)-4 \pi  \widetilde{\bar{\lambda}}^2 \tilde{\omega}_m \cos(2 \pi\tilde{\Phi}) \sinh (2\pi\tilde{\omega}_m)+\pi ^2 \tilde{\omega}_m^2\left[\cos(4 \pi\tilde{\Phi})+2\right]\\
\label{ISC_theta}
 &+&2 \cosh
   (2\pi\tilde{\omega}_m) \left[\widetilde{\bar{\lambda}}^4 \cos^2(\theta) \left(\cos (\varphi )-\cos(2 \pi\tilde{\Phi})\right)+2 \pi \widetilde{\bar{\lambda}}^2 \tilde{\omega}_m \sinh (2\pi\tilde{\omega}_m)-2 \pi ^2 \tilde{\omega}_m^2 \cos
   (2 \pi\tilde{\Phi})\right]\Big\}^{-1},
\end{eqnarray}
\end{widetext}
where  $\tilde{\omega}_m=\omega_m/\delta\varepsilon $ and $\widetilde{\bar{\lambda}} = \sqrt{2\pi} \,\bar{\lambda}/\hbar v_d$.
One can check that the kernel of the sum, and therefore the supercurrent, vanishes identically when $\theta=\frac{\pi}{2}$. This trivial case corresponds to the spins of the left and right Majorana being completely anti-parallel. 
\begin{figure}[t]
\includegraphics[width=\columnwidth]{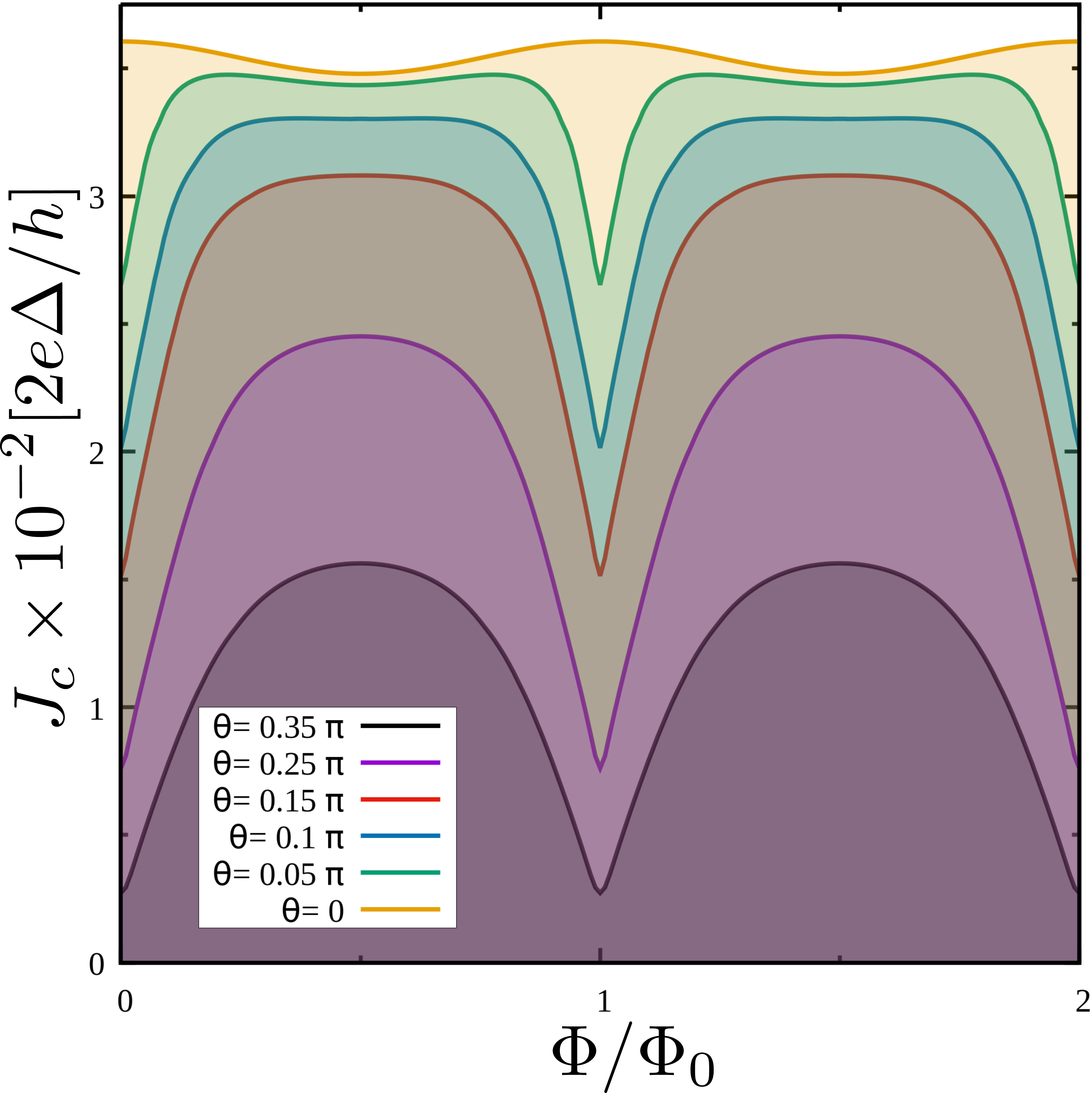}
\caption{Fraunhofer profiles obtained by maximizing Eq. (\ref{ISC_theta}) as function of the phase difference. Each curve is calculated for a different canting angle $\theta$. The parameters are such that $\delta\varepsilon =0.03$, $R=1$ and $\bar{\lambda}=0.02$. }
\label{fraunhofer_analytical}
\end{figure}

Fig. \ref{fraunhofer_analytical} shows the Fraunhofer patterns of this low energy model for different canting angles. These were numerically obtained by maximizing Eq. (\ref{ISC_theta}) as a function of $\varphi$ at zero temperature. We have chosen to normalize the critical current with the same units as in the tight-binding calculations, mainly $2e\Delta/h$ with $\Delta=0.3$, so as to properly compare the orders of magnitude. The qualitative behavior is very similar to the one obtained in the full tight-binding model of the junction: a series of dips arise when the electronic and hole modes become resonant at the Fermi level, an effect that within our model occurs at multiples of $\Phi_0$. As the canting angle decreases (which corresponds to an increase of $B_x$ in the leads) these dips tend to diminish their value with respect to the mean value of the critical current. In the Kitaev limit ($\theta=0$), these features completely disappear and the Fraunhofer profile turns into a smooth function of the flux variations.

Obtaining closed analytical expressions for the current-phase relation  [Eq.~\eqref{ISC_theta}] can be quite cumbersome. Nonetheless, for the particular cases where the flux threading the ring is an integer ($\tilde{\Phi}\in\mathbb{Z}$) or half an integer number ($\tilde{\Phi}\in\mathbb{Z}+1/2$) of flux quanta, some simplifications can be made. Even more, at zero temperature, the largest contribution to the supercurrent comes from the low frequency range and Eq.~\eqref{ISC_theta} can be roughly approximated by an integral of a Lorentzian shaped kernel. Within these estimates, we obtain that
\begin{eqnarray}
\notag
J^{N}_\mathrm{sc}(\varphi)&\simeq&
\frac{e\delta\varepsilon}{2h}\frac{\widetilde{\overline{\lambda}}}{\sqrt{2+\widetilde{\overline{\lambda}}^2}}\frac{\cos^2(\theta)\sin(\varphi)}{\sqrt{1-\cos^2(\theta)\sin^2(\varphi/2)}}\,,\\
J^{N+\frac{1}{2}}_\mathrm{sc}(\varphi) &\simeq& \frac{e\delta\varepsilon}{h}\frac{\widetilde{\overline{\lambda}}^2}{1+\widetilde{\overline{\lambda}}^2}\cos(\theta)\sin(\varphi/2)\text{sgn}(\pi-\varphi)\,,
\end{eqnarray}
where $J^{N}_\mathrm{sc}(\varphi)$ and $J^{N+\frac{1}{2}}_\mathrm{sc}(\varphi)$ are the approximated current-phase relations for an integer or half-integer number of flux quanta in the device, respectively. Note the sawtooth-like dependence of the supercurrent as a function of the phase difference $\varphi$ when there is a half-integer number of flux quanta threading the ring, capturing the topologically protected crossing between Andreev bound states at $\varphi=\pi$ [see Fig.~\ref{abs_analytical}(c)]. On the contrary, when $\tilde{\Phi}=N$ this functional form is smoothed by the canting angle $\theta$. This is due to the presence of a low energy gap between the $\omega=0$ flat band and the first dispersive Andreev bound state [see Fig. \ref{abs_analytical}(a)], which is essentially proportional to $\theta\delta\varepsilon/\pi$ when $\theta\rightarrow 0$. 

\begin{figure}[t]
\includegraphics[width=\columnwidth]{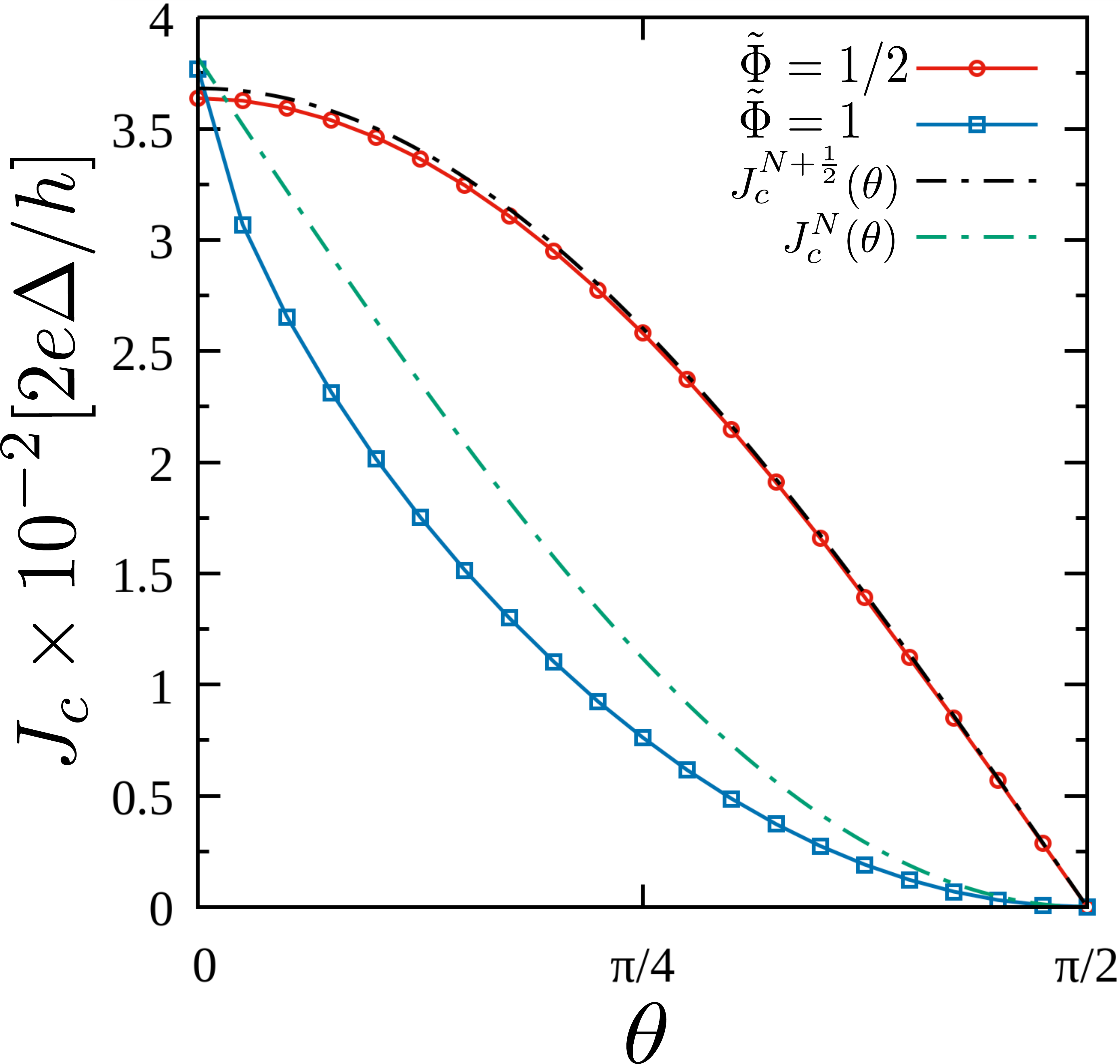}
\caption{Critical currents for $\tilde{\Phi}=1$ and $\tilde{\Phi}=1/2$ as a function of the canting angle $\theta$. Dashed lines are the approximate expressions for these corresponding magnitudes: $J^{N}_{c}(\theta)$ and $J^{N+\frac{1}{2}}_{sc}(\theta)$ (see Eq.~\eqref{Jc_approx}). The parameters are the same as the ones in Fig.~\ref{abs_analytical}.}
\label{critical_theta}
\end{figure}
The critical current for each of these scenarios is found to be
\begin{eqnarray}
\notag
J^{N}_{c}(\theta) &=& \frac{e\delta\varepsilon}{h}\frac{\widetilde{\overline{\lambda}}}{\sqrt{2+\widetilde{\overline{\lambda}}^2}}(1-\sin\theta)\,,\\
J^{N+\frac{1}{2}}_\mathrm{sc}(\theta) &=& \frac{e\delta\varepsilon}{h}\frac{\widetilde{\overline{\lambda}}^2}{1+\widetilde{\overline{\lambda}}^2}\cos\theta.
\label{Jc_approx}
\end{eqnarray}
\newline
\newline
Fig. \ref{critical_theta} shows the numerically obtained behavior of the critical current at zero temperature when $\tilde{\Phi}=1$ and $\tilde{\Phi}=1/2$ as a function of the canting angle $\theta$. The corresponding approximated analytical expressions given by Eq.~\eqref{Jc_approx} are shown in dashed lines. Even though these are not completely accurate, they are able to describe the general tendency: deep in the topological regime, when approaching $\theta=0$, the difference between $J_c(\tilde{\Phi}=1/2)$ and $J_c(\tilde{\Phi}=1)$ decreases, making the dips in the Fraunhofer pattern much less pronounced.

The high temperature limit  of Eq.~\eqref{ISC_theta} ($k_B T \gg \delta\varepsilon$) is given by
\begin{equation}
J_{sc}(\varphi) \simeq \frac{2 e \widetilde{\bar{\lambda}}^4}{\pi^4}\frac{\delta\varepsilon^2}{\hbar k_BT} \cos^2(\theta)\sin(\varphi)e^{-2\pi^2\frac{k_BT }{\delta\varepsilon}}.
\label{highT}
\end{equation}
In this regime, the supercurrent loses all the information on the flux threading the quantum Hall region because thermal effects wash out the level discretization of the edge state. However, the canting angle $\theta$ can be readily extracted from the critical current, since $J_c \propto \cos^2(\theta)$. We also note that the relevant length scale for the suppression of supercurrent is the perimeter of the sample, as expected for chiral edge mediated transport \cite{Ma1993,Stone2011,vanOstaay2011}. 

\subsection{Kitaev spinless limit}
When $\theta=0$ the physics of the device is exclusively determined by the $\hat{\psi}_1$ fields and the nanowires behave as Kitaev p-wave spinless chains. Taking this limit in Eq.~\eqref{ISC_theta}, we arrive to the following simplified expression for the supercurrent
\begin{widetext}
\begin{equation}
J_{sc}(\varphi) = \frac{e k_B T}{\hbar}\sum_m\frac{\widetilde{\overline{\lambda}}^4 \sin(\varphi)}{ \widetilde{\overline{\lambda}}^4\cos(\varphi) - \pi^2\tilde{\omega}_m^2\cos(2\pi\tilde{\Phi})+(\widetilde{\overline{\lambda}}^4 + \pi^2\tilde{\omega}_m^2)\cosh(2\pi\tilde{\omega}_m)+2\pi \widetilde{\overline{\lambda}}^2\tilde{\omega}_m\sinh(2\pi\tilde{\omega}_m)}\,.
\label{currentK}
\end{equation}
\end{widetext}
An alternative derivation of this expression is discussed in Appendix \ref{AppendixKitaev}. In Fig. \ref{kitaev} we show the critical current, obtained numerically from Eq. \eqref{currentK}  at zero temperature, as a function of the adimensional hopping amplitude $\widetilde{\overline{\lambda}}$. The two curves were calculated for $\tilde{\Phi}=1$ and $\tilde{\Phi}=1/2$. The inset shows the complete Fraunhofer interference patterns, each of them calculated for different magnitudes of $\widetilde{\overline{\lambda}}$. Notably, the critical current saturates for large hopping amplitudes and becomes independent of the variations of flux in the QH region. This behavior can be tracked down from the analytical expressions by realizing that, at zero temperature, the major contribution to the sum in Eq. \eqref{currentK} comes from the low frequency range. The supercurrent can then be approximated by
\begin{equation}
J_\mathrm{sc}(\varphi) \approx \frac{e}{2\pi\hbar}\delta\varepsilon\int_{-\infty}^{\infty} \frac{\widetilde{\overline{\lambda}}^4\sin(\varphi)d\tilde{\omega}}{2\widetilde{\overline{\lambda}}^4\cos^2(\varphi/2)+\pi^2\tilde{\omega}^2 X(\tilde{\Phi},\widetilde{\overline{\lambda}})}\,
\end{equation}
with 
 $X(\tilde{\Phi},\widetilde{\overline{\lambda}})=1-\cos(2\pi\tilde{\Phi})+2\widetilde{\overline{\lambda}}^4 + 4\widetilde{\overline{\lambda}}^2$.   
The integration is straightforward and we obtain
\begin{equation}
   J_\mathrm{sc}(\varphi)=J_c(\tilde{\Phi})\sin(\varphi/2)\text{sgn}(\pi-\varphi),
\label{JaproxK}
\end{equation}
where the critical current $J_c(\tilde{\Phi})$ is given by
\begin{equation}
J_c(\tilde{\Phi}) = \frac{e \delta\varepsilon}{\pi \hbar}\frac{\widetilde{\overline{\lambda}}^2}{\sqrt{2X(\tilde{\Phi},\widetilde{\overline{\lambda}})}}.
\end{equation}
Since in this calculations we implicitly assumed a thermodynamic average, Eq. \eqref{JaproxK} is $2\pi$-periodic in the phase difference $\varphi$ instead of $4\pi$-periodic. The fractional Josephson effect could be recovered by fixing the fermion parity, which would remove the sign function in the numerator. In any case, the expression for the critical current remains the same: it presents maximums whenever there is an integer number of normal flux quanta in the sample, a condition that makes the discrete levels of the QH region resonant with the Fermi level. 
\begin{figure}[t]
\includegraphics[width=\columnwidth]{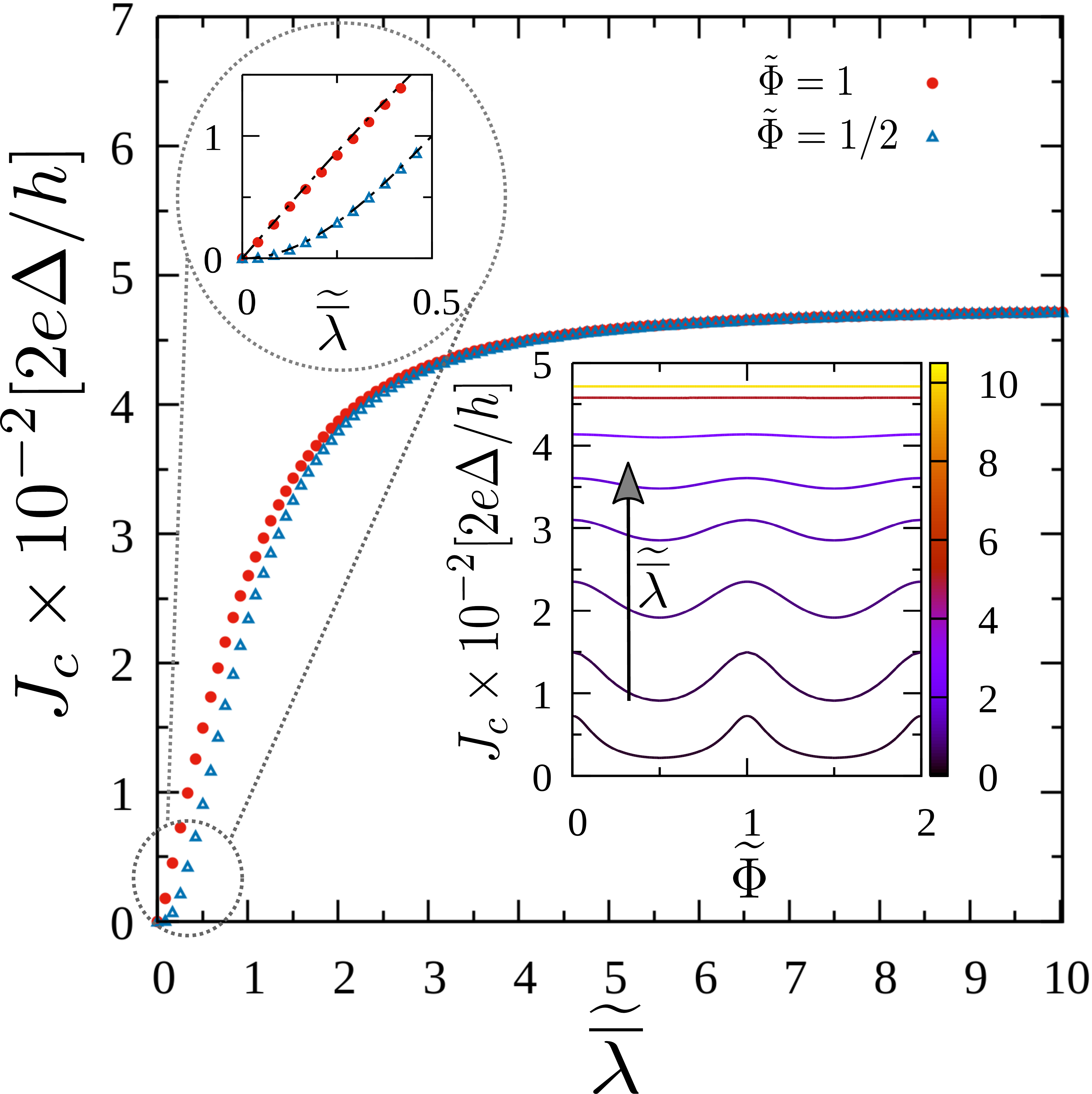}
\caption{Critical current $J_c$ in the Kitaev model ($\theta=0$) for $\tilde{\Phi}=1$ and $\tilde{\Phi}=1/2$ as a function of the adimensional hopping $\widetilde{\overline{\lambda}}$. These results were numerically obtained by maximizing Eq.~\eqref{currentK} as a function of $\varphi$ at zero temperature. The encircled inset is a zoom of the dependence for $\widetilde{\overline{\lambda}}\rightarrow 0$. The dashed lines were calculated with the respective approximations given by Eq.~\eqref{Jc_approx} when $\theta=0$. We also show the complete Fraunhofer patterns, where the color scale indicates the magnitude of $\widetilde{\overline{\lambda}}$. We have used the same parameters as in the main text, mainly $\delta\varepsilon=0.03$ and $R=1$.}
\label{kitaev}
\end{figure}

In the tunneling regime, when $\widetilde{\overline{\lambda}} \ll 1$, two different trends seem to appear. 
In the resonant case ($\tilde{\Phi}=N$ with $N\in \mathbb{Z}$), the critical current behaves as $J_c^{N} \simeq \frac{e\delta\varepsilon}{\sqrt{2} h}\widetilde{\overline{\lambda}}$. On the other hand, when the flux is detuned from this particular values, it switches from a linear dependence on the hopping amplitude to a quadratic one $J_c \simeq \frac{e\sqrt{2}\delta\varepsilon}{ h}\widetilde{\overline{\lambda}}^2\left[1-\cos(2\pi\tilde{\Phi})\right]^{-1/2}$. These behaviors are well captured by the full numerical integration of Eq.~\eqref{currentK}, as shown in the zoom of Fig.~\ref{kitaev}, where the dashed lines are the corresponding analytical expressions. We would like to emphasize that these linear and quadratic behaviors as a function of the hopping amplitude in the tunneling regime are characteristic of Majorana mediated transport through a resonant and off-resonant level, respectively. 

In the opposite limit, when $\widetilde{\overline{\lambda}}\gg1$, the dependence on the magnetic flux threading the QH is completely lost and the critical current tends to
\begin{equation}
\lim_{\widetilde{\overline{\lambda}}\to \infty}J_c = \frac{e}{h} \delta\varepsilon = e\frac{v_d}{2\pi R}.
\end{equation}
In this limit, the device behaves as a completely transparent long junction. The flux accumulated by an electron flowing from one lead to another is completely canceled out by the one of the perfectly Andreev reflected hole. This phenomenon is responsible for the aforementioned flux independence of the supercurrent. One can check that in this regime the Andreev bound states obtained from Eq.~\eqref{det} disperse linearly with the phase difference as $\omega_n =\pm \frac{\delta\varepsilon}{2\pi}(\varphi-\pi\pm n)$.
\section{Two edge channel transport results\label{IV}}
\begin{figure}[t]
\includegraphics[width=\columnwidth]{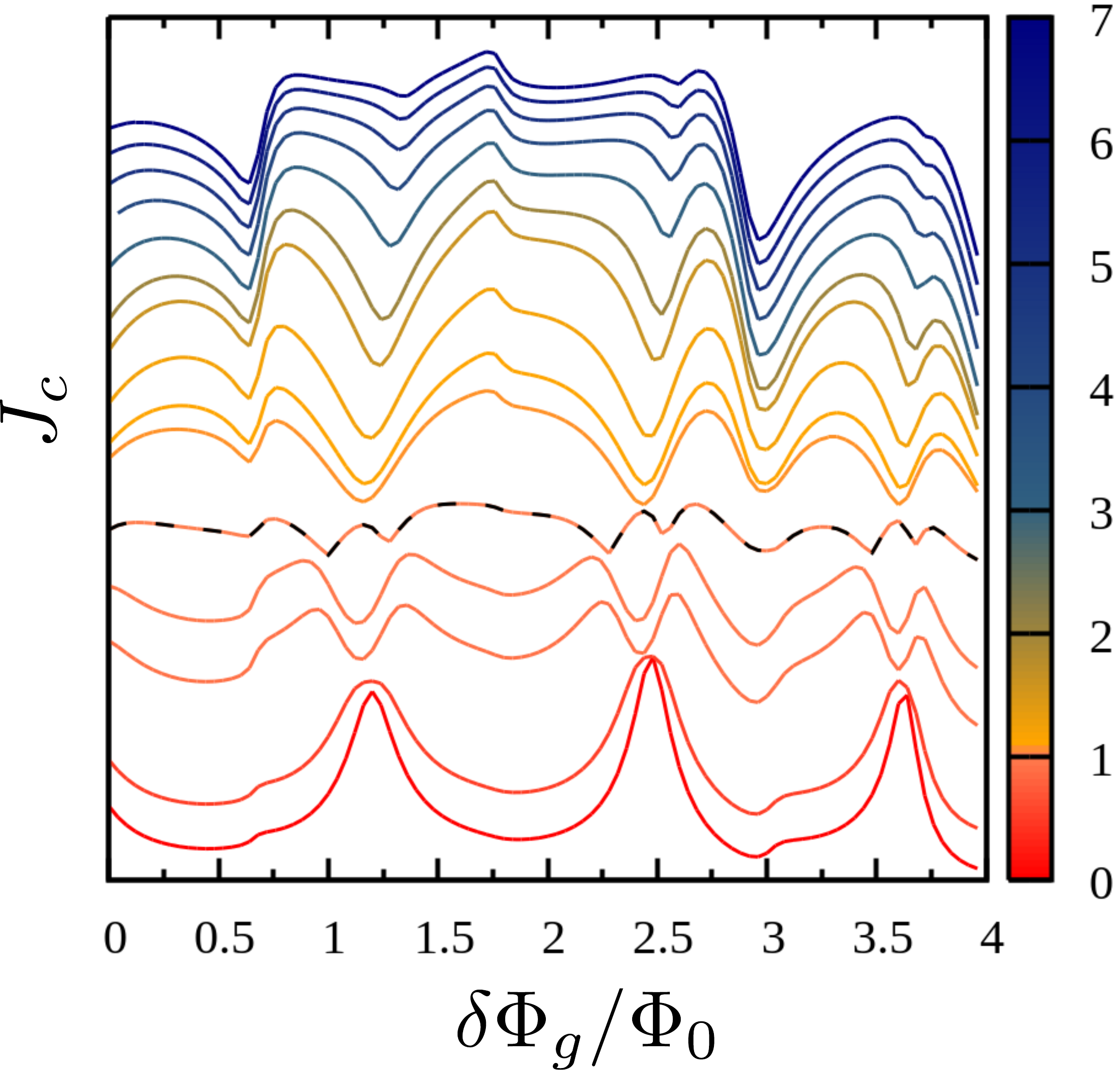}
\caption{Fraunhofer patterns: We show the critical current profiles as a function of the variations of total geometrical flux relative to the initial flux $\Phi_g$. The gate voltage in the Hall sample is chosen to be $V_g = 1.7$ so that two Landau levels are occupied. The color scale indicates the magnitude of the Zeeman field along the wires normalized to the critical field $B_x/B_c$. Dashed lines indicate the Fraunhofer pattern at the topological phase transition $B_x= B_c$. The curves are shifted for clarity.}
\label{fraunhofer2LL}
\end{figure}
We have so far concentrated on the case in which only the lowest Landau level was occupied. In this context, a simple one-dimensional model with electronic and hole chiral channels is enough to understand the basic physics of our results. Nonetheless, regimes where more than one Landau level is implicated are also experimentally accessible and of physical interest. In this case, the scenario becomes inherently more involved: each edge state can in principle interfere with the others in the Andreev reflection processes, all of them bearing different drift velocities and circulating along distinct effective perimeters. 

In this section we show how the tight-binding transport simulations change when the gate voltage in the Hall sample is chosen to be at $V_g = 1.7$, keeping all the other parameters the same. This choice ensures the occupation of two Landau levels in the QH region and already exhibits significant deviations with respect to our previous results. In Fig. \ref{fraunhofer2LL} we show the Fraunhofer interference patterns as a function of the variations of geometrical flux through the sample $\delta \Phi_g = \delta B_z A_g$ when modifying the Zeeman field $B_x$ along the wires. 
\begin{figure}[t]
\includegraphics[width=\columnwidth]{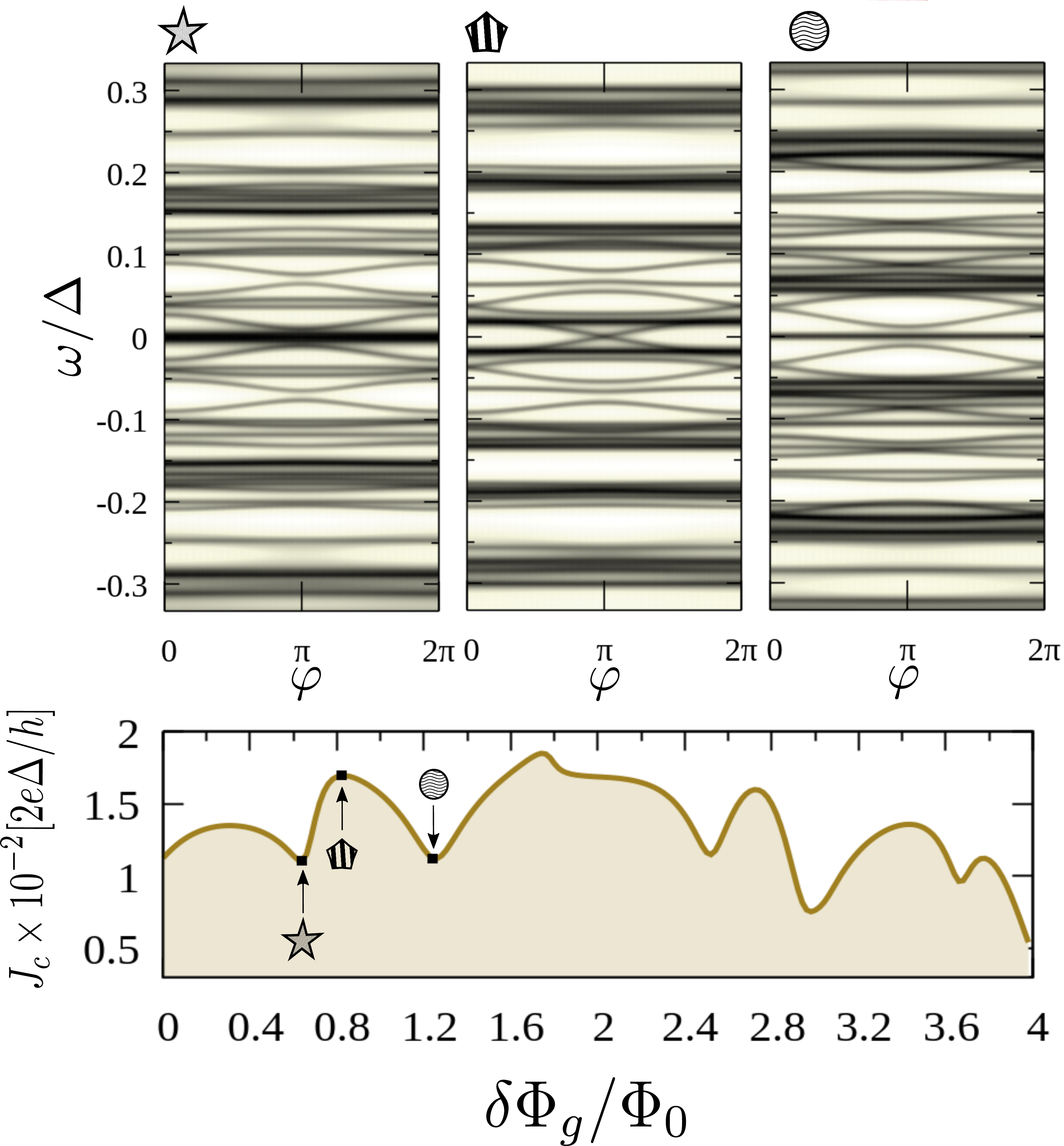}
\caption{The upper panels show the spectral density $\mathcal{A}_L(\omega,\varphi)$ in the quantum Hall regime with two occupied Landau levels. The color scale goes from white (zero) to black (higher value) in arbitrary units. Each figure is calculated for a different magnetic flux threading the central region, indicated by a corresponding symbol in the  Fraunhofer pattern shown in the lower panel. The nanowires are in the topological regime with a magnetic field $B_x = 2 B_c$.}
\label{abs2LL}
\end{figure}

The two sets of discretized levels coming from each edge channel generate a beating pattern with clearly more than one frequency involved. In general, the incommensurability of the spacing between the discrete levels arising from the first and second edge states and their mutual misalignment generates critical current profiles without a clear periodicity.

For the sake of completeness, we show in Fig. \ref{abs2LL} how the bound states of the system evolve for different fluxes when the Zeeman field is $B_x=2 B_c$. The chosen geometrical fluxes are marked with symbols in the corresponding Fraunhofer pattern shown in the lower panel. The presence of additional subgap states compared to the ones shown in Fig. \ref{absBx2-0} can be clearly identified. The dips in the critical current profiles are still correlated with the discrete levels becoming resonant at the Fermi level, but a complete understanding of these results requires a multi-channel analytical approach which is beyond the scope of the present work.

\section{Conclusions and final remarks}\label{V}
Throughout this work we have analyzed the transport and spectral properties of a quantum Hall based junction with superconducting leads that can be driven throughout a topological phase transition by tuning an external Zeeman field $B_x$. We have particularly focused on the case when only one Landau level is occupied, so there is a single chiral edge channel at the Fermi level. When the leads are in the trivial regime $B_x < B_c$, we recover some already known results: the Fraunhofer interference patterns obey a $\Phi_0$-periodicity when varying the flux threading the quantum Hall sample, a product of the existence of chiral edge states bridging the superconductors. This is manifested as a series of resonances in the critical current profiles that take place whenever the discrete levels that stem from the confinement of the edge channels along the perimeter of the Hall bar become aligned with the Fermi level. On the other hand, when the leads are in the topological regime $B_x > B_c$, the emergence of Majorana quasiparticles causes significant changes in these Fraunhofer modulations. The resonances become dips that possess a magnitude that is strongly dependent on the magnetic field along the nanowires. These results were understood within a low energy spinful model that allowed us to reproduce both the Andreev bound spectra and the Josephson current of the junction. The behavior of the spin polarization of the zero energy modes at the end sites of the one dimensional topological superconductors could be captured with the spin canting angle $\theta$, which has been shown to be responsible for the dip-like structure in the critical current profiles. We have also analyzed the $\theta=0$ limiting case, where the wires behave as Kitaev p-wave spinless chains. In this regime, closed analytical expressions for the Fraunhofer interference patterns could be extracted. We were able to pinpoint the pecularities of the Majorana mediated transport by analyzing the behavior of the critical current both in the tunneling and the strong-coupling regimes.

It is worth pointing out that our single-channel results  also apply to the case of a Hall bar made of graphene \cite{Amet2016,Seredinski2019,Zhao2020}, provided only the lowest Landau level is occupied and the Fermi energy is larger than the superconducting gap so the Dirac point physics \cite{Liu2017} is not involved. We have also checked that the addition of a small Zeeman term to the Hall bar Hamiltonian does not affect the Fraunhofer patterns provided the Landau levels of both spins are occupied and the Zeeman splitting is much smaller than the level spacing $\delta\varepsilon$, as assumed throughout the present work. In the case of graphene, this splitting is of the order of $0.1\,\text{meV}$ for magnetic fields of $B_z=1\,\text{T}$~\cite{Goerbig2011}, so that for samples with a perimeter of a few microns the level spacing is large enough.

The measurement of a supercurrent in this hybrid devices should be possible for sufficiently low temperatures. On one hand, the temperature should be small enough for the single-particle energy level spacing of the chiral edge modes to be resolved, otherwise the supercurrent is exponentially suppressed and the flux dependence lost [see Eq. \eqref{highT}]. Simple estimates for the sample used in Ref.~\cite{Amet2016} gives $\delta\varepsilon \approx 0.7\,\text{meV} = 800\,\text{mK}$, which is large compared with the temperatures of the order of $40-100\,\text{mK}$ that are used in typical transport experiments. On the other hand, the coherence length must be larger than the perimeter of the sample. Simple considerations in Ref.~\cite{Zhao2020} leads to $l_{\varphi}=\hbar vd/2 k_B T \approx 12\,\mu\text{m}$ for their graphene sample.

To conclude, we presented a detailed study of the evolution of the Fraunhofer oscillations in an integer quantum Hall sample when the superconducting leads are driven across a topological phase transition. 
Our results could be of relevance for
the detection of topological superconductivity and the general understanding of edge-channel transport of supercurrent in quantum Hall devices.

\acknowledgments
We acknowledge financial support from ANPCyT (grants PICTs 2016-0791 and 2018-01509 ), from CONICET (grant PIP 11220150100506) and from SeCyT-UNCuyo (grant 2019 06/C603). GU acknowledges support from the ICTP associateship program and thanks the Simons Foundation. LPG thanks R. Fazio for supporting her stay at the Condensed Matter Theory Group of the ICTP and Y. Gefen for fruitful discussions. 

\appendix
\section{Calculation of $\mathcal{G}(\alpha,\alpha,i\omega_m)$ \label{Green}}
We follow Ref.~[\onlinecite{Alavirad2018}] and choose a regularization scheme where
\begin{equation}
\mathcal{G}(\alpha,\alpha,i\omega_m) = \lim_{\epsilon\to 0}\frac{1}{2}\Big(\mathcal{G}(\alpha+\epsilon,\alpha,i\omega_m)+ \mathcal{G}(\alpha-\epsilon,\alpha,i\omega_m)\Big)\,.
\end{equation}
The matrices that propagate the Green's functions for $\alpha\,\in\,(0,\pi)$ and $\alpha\,\in\,(\pi,2\pi)$ are respectively given by
\begin{eqnarray}
\label{eqA}
\notag
\mathcal{G}(\alpha-\epsilon,\alpha,i\omega_m)_{\alpha\in(0,\pi)} &=& M_1(\alpha) \mathcal{G}(\alpha+\epsilon,\alpha,i\omega_m)_{\alpha\in(0,\pi)}\,,\\
\notag
\mathcal{G}(\alpha-\epsilon,\alpha,i\omega_m)_{\alpha\in(\pi,2\pi)} &=& M_2(\alpha) \mathcal{G}(\alpha+\epsilon,\alpha,i\omega_m)_{\alpha\in(\pi,2\pi)}\,,\\
\end{eqnarray}
where
\begin{eqnarray}
\nonumber
M_1(\alpha) &=& e^{-2\pi\tilde{\omega}_m}\mathcal{A}(\alpha)\mathcal{B}(\alpha)\tilde{M}_{0}\mathcal{A}(\pi)\mathcal{B}(-\pi)\\
&&\tilde{M}_{\pi}\mathcal{A}(\pi-\alpha)
\mathcal{B}(\pi-\alpha)\\
\notag
M_2(\alpha) &=& e^{-2\pi\tilde{\omega}_m}\mathcal{A}(\alpha-\pi)\mathcal{B}(\pi-\alpha)\tilde{M}_{\pi}\mathcal{A}(\pi)\mathcal{B}(\pi)\\
\notag
&&\tilde{M}_0\mathcal{A}(2\pi-\alpha)\mathcal{B}(\alpha-2\pi)\,,
\end{eqnarray}
and 
\begin{equation}
\mathcal{A}(x)=e^{i x\tilde{\Phi}\sigma_0\otimes\tau_z}\,,\qquad \mathcal{B}(x)=e^{-ix\frac{\varphi}{4\pi}(\sigma_0+\sigma_z)\otimes\tau_z}\,,
\end{equation} 
with $\tilde{\omega}_m = \omega_m/\delta\varepsilon$.
Here, the transfer matrices $\tilde{M}_0 = M_0(i\omega_m,0)$ and $\tilde{M}_{\pi} = M_{\pi}(i\omega_m)$ no longer depend on the superconducting phase difference between the leads $\varphi$, since it has been incorporated as a vector potential in the propagators. 
On the other hand, for angles belonging to the intervals $(0,\pi)$ and $(\pi,2\pi)$, we can integrate the Dyson equation in $\alpha$ to obtain the relation
\begin{equation}
i\hbar v_d\Big(\mathcal{G}(\alpha+\varepsilon,\alpha,i\omega_m)-\mathcal{G}(\alpha-\varepsilon,\alpha,i\omega_m)\Big)=\mathbb{1}\,,
\label{eqB}
\end{equation}
where $\mathbb{1}$ is the $4\times4$ identity matrix.
With this information [Eqs. (\ref{eqA}) and (\ref{eqB})] we can now write the local Green's functions in terms of the $M_1$ and $M_2$ as
\begin{eqnarray}
\mathcal{G}(\alpha,\alpha,i\omega_m)_{\alpha\in(0,\pi)} &=&\frac{-i}{2\hbar v_d}(\mathbb{1}+M_1)(\mathbb{1}-M_1)^{-1}\,,\\
\notag
\mathcal{G}(\alpha,\alpha,i\omega_m)_{\alpha\in(\pi,2\pi)} &=&\frac{-i}{2\hbar v_d}(\mathbb{1}+M_2)(\mathbb{1}-M_2)^{-1}\,.
\end{eqnarray}
When replacing these expressions in Eq.~\eqref{Jsc_G} in the main text, the supercurrent takes the form
\begin{eqnarray}
\notag
J_\text{sc} &=&-\frac{i e k_BT} {8\hbar}\sum_{m}\Big\{\text{Tr}\Big[(\sigma_0+\sigma_z)\otimes\tau_z (\mathbb{1}+M_1)(\mathbb{1}-M_1)^{-1}\Big]\\
&&-\text{Tr}\Big[(\sigma_0+\sigma_z)\otimes\tau_z (\mathbb{1}+M_2)(\mathbb{1}-M_2)^{-1}\Big]\Big\},
\label{Jsc_M1M2}
\end{eqnarray}
where we have used the fact that the traces are independent of the angle $\alpha$.
\section{ Kitaev limit within a Green's function approach.\label{AppendixKitaev}}

In this appendix we introduce yet another approach for the derivation of the supercurrent in this quantum Hall device. Our purpose is to present an alternative description of the transport properties of the junction within a Green's function formalism, instead of the scattering technique used in Section \ref{III}. We analyze in particular the Kitaev limit---which corresponds to the limiting case of $\theta=0$ in the model depicted in Fig. \ref{modelfig}---where the leads are considered as spinless one-dimensional p-wave superconductors. 

The low-energy Hamiltonian describing this setup is given by
\begin{eqnarray}
\notag
H_{K}&=& H_{qh}-\frac{t}{2} \sum_n\hat{\gamma}_{R}^{}\left(e^{-i \varphi /2}\hat{c}_{n}^{} -e^{i \varphi /2}\hat{c}^{\dagger}_{n}\right)\\
& &-\frac{t}{2}i \sum_n\hat{\gamma}_{L}^{}\left(e^{i\pi n}\hat{c}_{n}^{} +e^{-i\pi n}\hat{c}^{\dagger}_{n}\right).
\end{eqnarray}
Here, $H_{qh}$ describes the QH central region of spinless fermions, and $\hat{\gamma}_{R}$ and $\hat{\gamma}_{L}$ are Majorana operators acting at the edges of the right and left superconducting wires, respectively. The operator $\hat{c}^{\dagger}_n$ ($\hat{c}_n^{}$) creates (destroys) a particle in an eigenstate of the uncoupled ring.

The current flowing from the left contact to the QH region is then expressed as 
\begin{equation}
\langle \hat{J}_{sc}\rangle\!=\! \frac{2e}{\hbar}\left\langle\frac{\partial H_K}{\partial \varphi}\right\rangle\!= \frac{i te}{2\hbar}\sum_n\left[e^{i\varphi/2}\langle\hat{\gamma}_R\hat{c}^{\dagger}_n\rangle - e^{-i\varphi/2}\langle\hat{c}_n^{}\hat{\gamma}_R\rangle \right].
\end{equation}
At finite temperature $T$, these mean values can be written in terms of the Matsubara Green's function between the right Majorana and the $n$-th state, with the fermionic Matsubara frequency defined as $\omega_m = (2m +1)\pi k_B T$. By means of the equations of motion, all the one-particle Green's functions can be obtained, and after some algebra the current is found to be
\begin{equation}
J_{sc}= \frac{2e k_B T}{\hslash}\sum_{m}\sin(\varphi)
 A_{J}(i\omega_m, \varphi),
\label{Jkitaev}
\end{equation}
with
\begin{equation}
A_{J}(\omega, \varphi)= \frac{(t^2/2\omega)^2\mathcal{G}^{-}_{0\pi}(\omega)\mathcal{G}^{+}_{0\pi}(\omega)}{[1-D(\omega,\varphi)][1-D_0(\omega)]}.
\label{AJ}
\end{equation}
Here we have defined
\begin{equation}
D_0(\omega)= \frac{t^2}{2 \omega}\left[\mathcal{G}^{-}_{00}(\omega)+\mathcal{G}^{+}_{00}(\omega)\right],
\end{equation}
and
\begin{eqnarray}
D(\omega, \varphi)&=&D_0(\omega)+\left(\frac{t^2}{2 \omega}\right)^2\frac{1}{1-D_0(\omega)}\times\\
\nonumber
&&[\mathcal{G}^{-}_{\pi 0}(\omega)- e^{i\varphi}\mathcal{G}^{+}_{0\pi}(\omega)][\mathcal{G}^{-}_{0\pi}(\omega)- e^{-i\varphi}\mathcal{G}^{+}_{\pi 0}(\omega)],
\end{eqnarray}
where $\mathcal{G}^{\mp}_{\alpha\beta}(\omega)$ are the electron ($-$) and hole ($+$) propagators of the central QH region. Note the presence of the pair susceptibility of the device $\mathcal{G}^{-}_{0\pi}(\omega)\mathcal{G}^{+}_{0\pi}(\omega)$ in Eq. \eqref{AJ}, which reveals the propagation of an electron and a hole from the site located at angle $\pi$ to the one at angle $0$. The numerator in Eq. \eqref{AJ} actually bears a resemblance with the perturbative findings of Ref.~\citep{Ma1993}, but where the BCS superconductors Green's function has been replaced by the Majorana singularity at zero energy. 

For the particular case of the extreme quantum limit, where only one Landau Level is occupied, these propagators acquire a simple form. By making use of the Lehmann spectral representation, the diagonal propagators turn out to be
\begin{eqnarray}
\nonumber
\mathcal{G}^{\mp}_{\pi\pi}(\omega)=\mathcal{G}^{\mp}_{00}(\omega)=\frac{1}{\delta\varepsilon} \sum_{n} \frac{1}{\tilde{\omega}\mp n \pm \tilde{\Phi}}=\frac{\pi}{\delta\varepsilon } \cot(\pi(\tilde{\omega} \pm\tilde{\Phi}))\,,\\
\end{eqnarray}
where we made use of the notation of Section \ref{III} by writing the eigenvalues of the central region as  $E_n=\delta\varepsilon (n-\tilde{\Phi})$ and defined $\tilde{\omega}=\omega/\delta\varepsilon$. The Fermi level has been taken to be zero for simplicity. Similarly, the non-diagonal propagators are
\begin{eqnarray}
\nonumber
\mathcal{G}^{\mp}_{\pi 0}(\omega)=\mathcal{G}^{\mp}_{0\pi}(\omega)=\frac{1}{ \delta\varepsilon} \sum_{n} \frac{e^{in\pi}}{\tilde{\omega}\mp n \pm \tilde{\Phi}}=\frac{\pi}{\delta\varepsilon } \csc(\pi(\tilde{\omega} \pm\tilde{\Phi})).\\
\end{eqnarray}

One can check that these expressions reproduce Eq. \eqref{currentK} in the main text by replacing Eq. \eqref{AJ} in Eq. \eqref{Jkitaev} and taking $\frac{t}{2} = \frac{\overline{\lambda}}{\sqrt{2\pi}R} = \frac{\delta\varepsilon\widetilde{\overline{\lambda}}}{2\pi}$.

\end{document}